\documentclass{frontiersFPHY}

\usepackage{graphicx,color,rotating,pifont}
\usepackage{amsmath,amssymb,bm,bbm}
\usepackage{ae}
\usepackage{pstricks}
\usepackage{simplewick}

\setcitestyle{square} 
\usepackage{url,hyperref,lineno,microtype,subcaption}
\usepackage{subcaption}
\usepackage[onehalfspacing]{setspace}
\usepackage{longtable}
\allowdisplaybreaks

\renewcommand{\vec}[1]{\ensuremath{\boldsymbol{#1}}}

\DeclareMathSymbol{\NS}{\mathord}{AMSb}{"4E}

\newcommand{\bra}[1]{\ensuremath{\langle{#1}|\,}}
\newcommand{\ket}[1]{\ensuremath{\,|{#1}\rangle}}
\newcommand{\braket}[2]{\ensuremath{\langle{#1}|{#2}\rangle}}
\newcommand{\matrixe}[3]{\ensuremath{\langle{#1}|\,{#2}\,|{#3}\rangle}}

\newcommand{\braketn}[1]{\ensuremath{\braket{#1}{#1}}}

\newcommand{\dmatrixe}[2]{\matrixe{#1}{#2}{#1}}

\newcommand{\expect}[1]{\ensuremath{\langle{#1}\rangle}}

\newcommand{\comm}[2]{\ensuremath{[{#1},{#2}]}}

\newcommand{\op}[1]{\ensuremath{#1}}
\newcommand{\adj}[1]{\ensuremath{{{#1}}^{\dagger}}}

\newcommand{\totd}[2]{\ensuremath{ \frac{d {#1}} {d {#2}} }}

\newcommand{\aO}{\ensuremath{\op{a}}}

\newcommand{\etaO}{\ensuremath{\op{\eta}}}

\newcommand{\aaO}{\ensuremath{\adj{\op{a}}}}

\newcommand{\HO}{\ensuremath{\op{H}}}

\newcommand{\OO}{\ensuremath{\op{O}}}

\newcommand{\UO}{\ensuremath{\op{U}}}

\newcommand{\UUO}{\ensuremath{\adj{\op{U}}}}

\newcommand{\idO}{\ensuremath{\op{1}}}

\newcommand{\nuc}[2]{\ensuremath{^{#2}\mathrm{#1}}}

\newcommand{\fmi}{\ensuremath{\,\text{fm}^{-1}}}

\newcommand{\keV}{\ensuremath{\,\text{keV}}}
\newcommand{\MeV}{\ensuremath{\,\text{MeV}}}

\newcommand{\NNNNLO}{N$^4$LO}
\newcommand{\NNNLO}{N$^3$LO}
\newcommand{\NNLO}{NNLO}
\newcommand{\NLO}{NLO}
\newcommand{\NNLOsat}{$\text{NNLO}_\text{sat}$}
\newcommand{\lambdaSRG}{\ensuremath{\lambda}}

\definecolor{FGViolet}{rgb}{0.61,0.32,0.61}
\definecolor{FGDarkBlue}{rgb}{0,0,0.6}
\definecolor{FGBlue}{rgb}{0,0,0.8}
\definecolor{FGLightBlue}{rgb}{0.2, 0.6, 0.8}
\definecolor{FGGreen}{rgb}{0.2,0.7,0.2}
\definecolor{FGLightGreen}{rgb}{0.4,1,0.4}
\definecolor{FGYellow}{rgb}{1,0.95,0}
\definecolor{FGOrange}{rgb}{0.95,0.5,0.1}
\definecolor{FGRed}{rgb}{0.8,0,0}
\definecolor{FGWhite}{rgb}{1,1,1}
\definecolor{FGLightGray}{rgb}{0.8,0.8,0.8}
\definecolor{FGGray}{rgb}{0.5,0.5,0.5}
\definecolor{FGDarkGray}{rgb}{0.3,0.3,0.3}
\definecolor{FGBlack}{rgb}{0,0,0}

\def\keyFont{\fontsize{8}{11}\helveticabold }
\def\firstAuthorLast{Hergert} 
\def\Authors{Heiko Hergert\,$^{1,*}$}
\def\Address{$^{1}$Facility for Rare Isotope Beams and Department of Physics \& Astronomy, Michigan State University,
East Lansing, MI 48824-1321}

\def\corrAuthor{Heiko Hergert}
\def\corrEmail{hergert@frib.msu.edu}

\extraAuth{}

\begin{document}
\onecolumn
\firstpage{1}

\title[A Guided Tour of \emph{Ab Initio} Nuclear Many-Body Theory]{A Guided Tour of \emph{Ab Initio} Nuclear Many-Body Theory} 

\author[\firstAuthorLast ]{\Authors} 
\address{} 
\correspondance{} 

\maketitle

\begin{abstract}

Over the last decade, new developments in Similarity Renormalization 
Group techniques and nuclear many-body methods have dramatically increased 
the capabilities of \emph{ab initio} nuclear structure and reaction theory.
Ground and excited-state properties can be computed up to the tin region,
and from the proton to the presumptive neutron drip lines, providing unprecedented
opportunities to confront two- plus three-nucleon interactions from chiral 
Effective Field Theory with experimental data. In this contribution, I will
give a broad survey of the current status of nuclear many-body approaches,
and I will use selected results to discuss both achievements and open issues 
that need to be addressed in the coming decade.

\tiny
 \keyFont{ \section{Keywords:} nuclear theory, many-body theory, \emph{ab initio} nuclear structure,  \emph{ab initio} nuclear reactions, similarity renormalization group} 
\end{abstract}

\section{Introduction}
\label{sec:intro}

Over the past decade, the reach and capabilities of \emph{ab initio} nuclear 
many-body theory have grown exponentially. The widespread adoption of Renormalization 
Group (RG) techniques, in particular the Similarity Renormalization Group 
(SRG) \cite{Bogner:2010pq}, and Effective Field Theory (EFT) 
\cite{Machleidt:2016yo,Epelbaum:2020rr,Piarulli:2020dp} in the 2000s laid 
the foundation for these developments. Consistent two-nucleon (NN) and three-nucleon 
(3N) interactions from chiral EFT were quickly established as a new 
``standard'' inputs for a variety of approaches, which made true multi-method 
benchmarks possible. The SRG equipped us with the ability to dial the 
resolution scale of nuclear interactions, accelerating model-space and 
many-body convergence alike. Suddenly, even (high-order) Many-Body 
Perturbation Theory (MBPT) became a viable tool for rapid benchmarking 
\cite{Roth:2010ys,Tichai:2016vl}, and exact diagonalization approaches were 
able to extend their reach into the lower $sd$-shell 
\cite{Navratil:2007fk,Roth:2007fk,Roth:2009eu}. A variety of of computationally 
efficient techniques with controlled truncations were readied, like the 
Self-Consistent Green's Function method (SCGF) \cite{Soma:2020fj}, the 
In-Medium SRG (IMSRG) \cite{Hergert:2016jk} and Coupled Cluster (CC) 
\cite{Hagen:2014ve}, the prodigal son \cite{Coester:1958dq,Coester:1960cr} 
who returned home after finding success in foreign lands, i.e., quantum 
chemistry and solid state physics. 

\begin{figure}[t]
  \begin{center}
    \setlength{\unitlength}{\textwidth}
    \includegraphics[width=\unitlength]{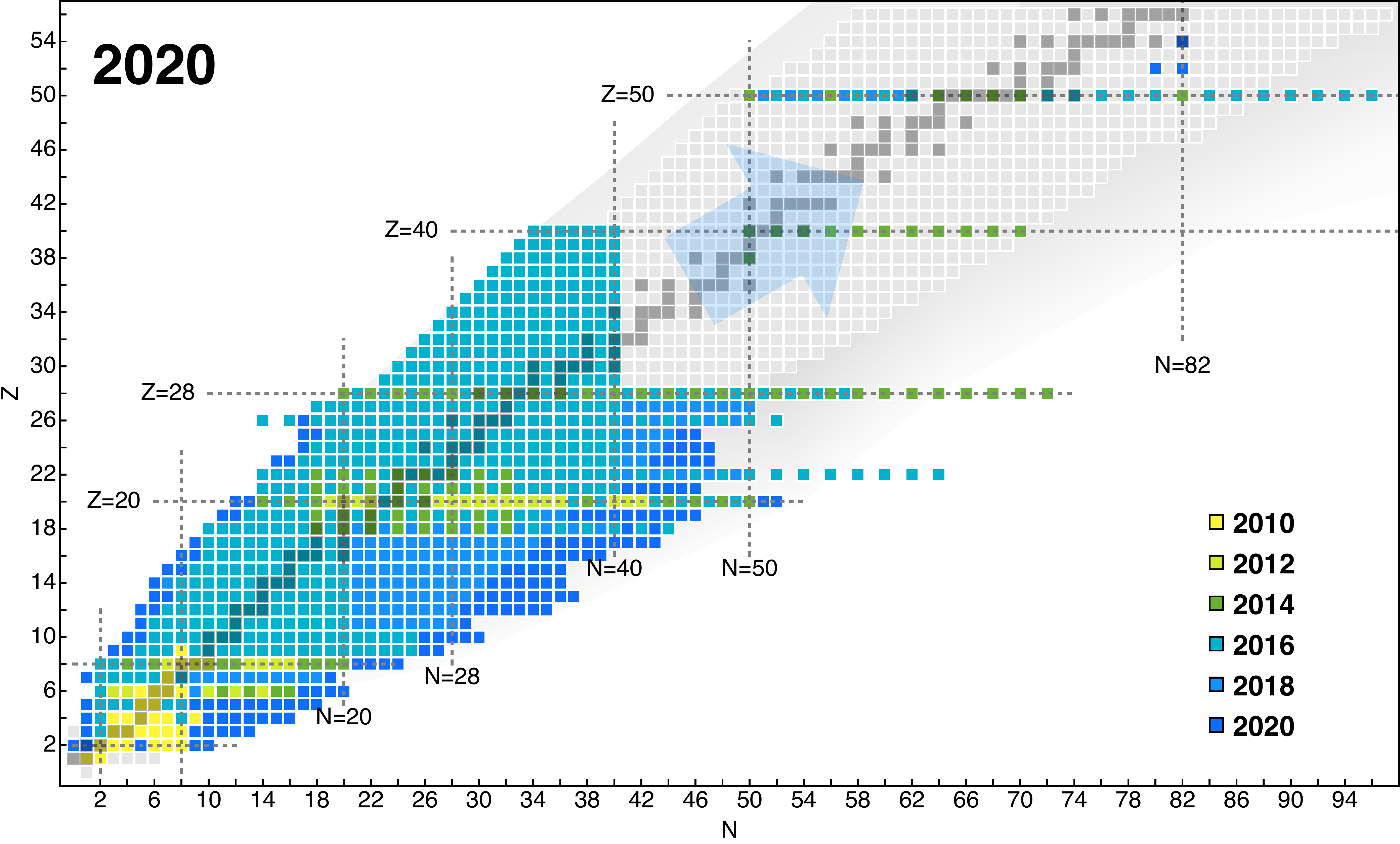}
  \end{center}
  \vspace{-5pt}
  \caption{
    Progress in \emph{ab initio} nuclear structure calculations over the past decade.
    The blue arrow indicates nuclei that will become accessible with new advances
    for open-shell nuclei in the very near term (see Sec.~\ref{sec:many_body}).
  }
  \label{fig:nuclear chart}
\end{figure}

At the start of the last decade the race was on, and Fig.~\ref{fig:nuclear chart}
documents the progress that ensued. Calculations started at closed-shell 
nuclei \cite{Hagen:2012oq,Hagen:2012nx,Hergert:2013mi,Cipollone:2013uq,Binder:2014fk} 
and their vicinity before extending to semi-magic isotopic chains 
with the development of the Multi-Reference IMSRG \cite{Hergert:2013ij,Hergert:2014vn} 
and Gor'kov SCGF \cite{Soma:2013ys,Soma:2014eu} techniques, and just a couple 
of years later, the use of CC \cite{Jansen:2014qf,Sun:2018ul} and 
IMSRG \cite{Bogner:2014tg,Stroberg:2017sf} techniques to construct valence-space 
interactions opened all nuclei that were amenable to Shell Model calculations 
for exploration. Owing to very recent developments that extend these combined 
approaches to multi-shell valence spaces, the open region between the nickel 
and tin isotopic chain is poised to be filled in rapidly \cite{Miyagi:2020rm}. 
Development of the no-core versions of these methods has continued as well, 
and made direct calculations for intrinsically deformed nuclei possible 
\cite{Yao:2020mw}. 

The growing reach of \emph{ab initio} many-body methods made it possible to 
confront chiral NN+3N forces with a wealth of experimental data, revealing
shortcomings of those interactions and sparking new efforts toward their 
improvement. There were other surprises along the way, some good, some
bad. Due to the benchmarking capabilities and further developments in many-body
theory, we are now often able to understand the reasons for the failure of 
certain calculations (see, e.g., Ref.~\cite{Stroberg:2017sf}) --- hindsight 
is 2020, as they say\footnote{This exhausts my contractually allowed contingent 
of 2020 vision puns, I swear.}. 

The present collection of Frontiers in Physics contributions provides us with 
a timely and welcome opportunity to attempt a look back at some of the impressive 
results from the past decade and the developments that brought us here, as well 
as a look ahead at the challenges to come as we enter a new decade.

Let us conclude this section with a brief outline of the main body of this 
work. In Section \ref{sec:ingredients}, I will discuss the main ingredients 
of modern nuclear many-body calculations: The input interactions from chiral EFT, the 
application of the SRG to process Hamiltonians and operators, and eventually 
a variety of many-body methods that are used to solve the Schr\"{o}dinger 
equation. I will review key ideas but keep technical details to a minimum, 
touching only upon aspects that will become relevant again later on. 
Section \ref{sec:current} presents selected applications from the past decade, 
and discusses both the advances they represent as well as open issues. This 
will provide a starting point for Section \ref{sec:future}, which presents 
ideas for addressing the aforementioned issues and highlights important 
directions for the next decade.

Naturally, the discussion in Sections \ref{sec:current} and \ref{sec:future} 
is highly subjective. While this work grew from a more restricted scope into 
a rambling, albeit not random, walk through the landscape of modern nuclear 
many-body theory, it still cannot encompass the field in its entirety. The 
upside is that this reflects the breadth of ideas that are being pursued by 
the \emph{ab initio} nuclear theory community, including those with 
cross-disciplinary impact, as well as our community's ability to attract 
junior researchers. The downside is that the present work can only scratch 
the tip of the iceberg of impressive results from the past decade. I hope 
that the readers will use it as a jumping-off point for delving into the 
cited literature, including the contributions to this volume. 

\section{Players on a Stage: Elements of Nuclear Many-Body Theory}
\label{sec:ingredients}

\subsection{Interactions from Chiral Effective Field Theory}
\label{sec:eft}
Quantum Chromodynamics (QCD) is the fundamental theory of the strong interaction
between quarks and gluons. One of its characteristic features is that the strong 
coupling, which governs the strength of interaction processes, is sufficiently 
small to allow perturbative expansions at high energies, but large in the low-energy 
domain relevant for nuclear structure and dynamics \cite{Gross:1973pd,Politzer:1973lq}. 
This makes the description of all but the lightest nuclei at the QCD level inefficient 
at best, and impossible at worst. However, strongly interacting matter undergoes a 
phase transition that leads to the confinement of quarks in composite hadronic 
particles, like nucleons and pions. These particles can be used as the degrees of
freedom for a hierarchy of EFTs that describe the strong interaction across multiple
scales.

\begin{figure}[t]
  \setlength{\unitlength}{0.8\textwidth}
  \begin{center}
    \includegraphics[width=\unitlength]{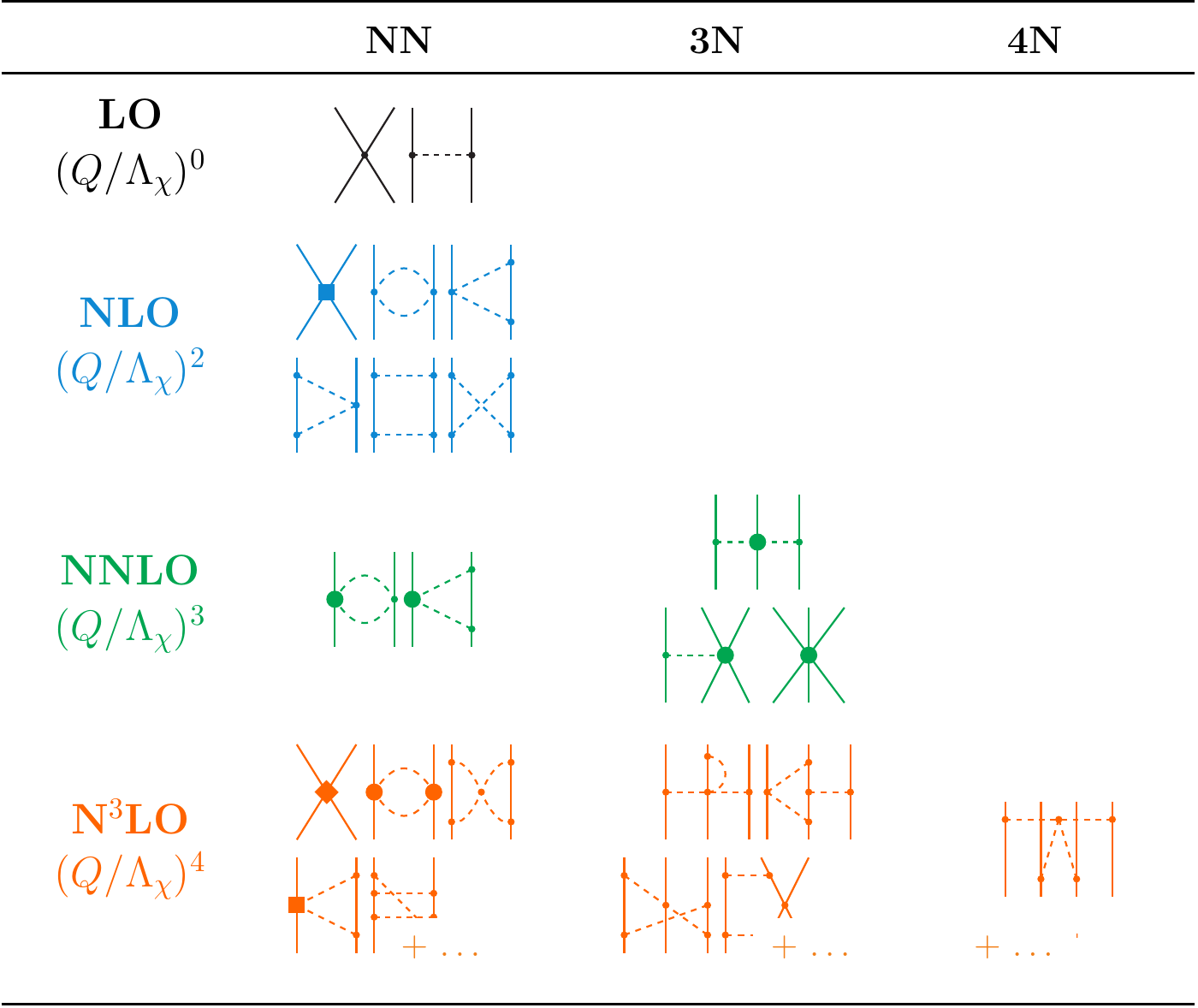}
  \end{center}
  \caption{Chiral two-, three- and four nucleon forces 
  through next-to-next-to-next-to-leading order (\NNNLO{}) (see, e.g., 
  \cite{Epelbaum:2009ve,Machleidt:2011bh,Machleidt:2016yo}
  ). Dashed lines
  represent pion exchanges between nucleons. The large solid circles, boxes
  and diamonds represent vertices that are proportional to low-energy constants (LECs) of
  the theory (see text). 
  }
  \label{fig:chiral_forces}
\end{figure}

Following Weinberg \cite{Weinberg:1991rw,Weinberg:1996bb}, one can construct
effective Lagrangians that consist of interactions that are consistent with the 
symmetries of QCD and organized by an expansion in ($Q/\Lambda$). Here, $Q$ 
is a typical momentum of the interacting system, and $\Lambda$ is the breakdown
scale of the theory, which is associated with physics that is not explicitly
resolved. In chiral EFT with explicit nucleons and pions, $\Lambda = \Lambda_\chi$ 
is traditionally considered to be in the range $700-1000\,\MeV$, although
newer analyses of observable truncation errors using Bayesian methods
favor slightly lower values \cite{Melendez:2017fj,Melendez:2019ax,Wesolowski:2019zv}.
From a chiral EFT Lagrangian, one can then construct a systematic low-momentum 
expansion of nuclear interactions, as shown in Fig.~\ref{fig:chiral_forces} 
(see Refs.~\cite{Weinberg:1991rw,Epelbaum:2009ve,Machleidt:2016yo,Epelbaum:2020rr,Rodriguez-Entem:2020oq}).
These interactions consist of (multi-)pion exchanges between nucleons, indicated
by dashed lines, as well as nucleon contact interactions. The different types of
vertices are proportional to the low-energy constants (LECs) of chiral EFT,
which encode physics that is not explicitly resolved because it involves either a 
high momentum scale or excluded degrees of freedom. Eventually, one hopes to 
calculate these LECs directly from the underlying QCD either through matching
or renormalization group evolution of the couplings (see Section \ref{sec:srg}), 
but at present, the LECs are fit to experimental data \cite{Epelbaum:2020rr,Rodriguez-Entem:2020oq,
Piarulli:2020dp,Ekstrom:2020la,Ruiz-Arriola:2020pi}.

The power counting scheme shown in Fig.~\ref{fig:chiral_forces} yields consistent
two-, three- and higher many-nucleon interactions, and explains their empirical 
hierarchy, i.e., $V_\text{NN}>V_\text{3N}>V_\text{4N}>\ldots$. Moreover, one can
readily extend the chiral Lagrangian with couplings to the electroweak sector by 
gauging the derivatives. In this way, nuclear interactions and electroweak currents
depend on the same LECs, and one can use electroweak observables to constrain their
values \cite{Gazit:2009qf,Pastore:2011dq,Kolling:2011bh,Piarulli:2013vn}.
Last but not least, the existence of a power counting scheme offers inherent
diagnostics for assessing the theoretical uncertainties that result from working
at a given chiral order \cite{Melendez:2017fj,Melendez:2019ax,Wesolowski:2019zv}.
This is especially useful since issues relating to the regularization and renormalization 
of these interactions remain (see, e.g., Refs.~  
\cite{Machleidt:2016yo,Reinert:2018uq,Lynn:2016ec,Lynn:2017eu,Valderrama:2017hb,Sanchez:2018fr,Kolck:2020dn}
and Sec.~\ref{sec:eft_future}).

\subsection{The Similarity Renormalization Group}
\label{sec:srg}
Renormalization group methods are a natural companion to the hierarchy
of EFTs for the strong interaction. They provide the means to systematically 
dial the resolution scales and cutoffs of these theories, and this makes it 
possible, at least in principle, to connect the different levels in our 
hierarchy of EFTs. The RGs also \emph{expand} the diagnostic toolkit
for assessing the inherent consistency of EFT power counting schemes, e.g., by 
tracing the enhancement or suppression of specific operators, or by identifying 
important missing operators.

In nuclear many-body theory, the SRG has become the method of choice. In 
contrast to Wilsonian RG \cite{Wilson:1975qo}, which is based on \emph{decimation}, 
i.e., integrating out high-momentum degrees of freedom, SRGs decouple low- 
and high-momentum physics using continuous unitary transformations. Note 
that this concept is not limited to RG applications: we can construct 
transformations that adapt a many-body Hamiltonian or other observables 
of interest to our needs, e.g., to extract eigenvalues \cite{Hergert:2016jk,Hergert:2017kx},
or impose specific structures on the operator 
\cite{Bogner:2010pq,Bogner:2014tg,Stroberg:2016fk,Stroberg:2017sf,Johnson:2020kk}.

We define the flowing Hamiltonian
\begin{equation}\label{eq:cut}
  \HO(s)=\UO(s)\HO(0)\UUO(s)\,,
\end{equation}
where $H(s=0)$ is the starting Hamiltonian, and the flow parameter $s$ parameterizes 
the unitary transformation. Instead of making an ansatz for $U(s)$, we take the derivative 
of Eq.~\eqref{eq:cut} and obtain the operator flow equation
\begin{equation}\label{eq:opflow}
  \totd{}{s}\HO(s) = \comm{\etaO(s)}{\HO(s)}\,,
\end{equation} 
where the anti-Hermitian generator $\etaO(s)$ is related to $\UO(s)$ by
\begin{equation}
  \eta(s)=\totd{U(s)}{s}U^{\dagger}(s) = -\eta^{\dagger}(s)\,.
\end{equation}
We can choose $\etaO(s)$ to achieve the desired transformation of the Hamiltonian
as we integrate the flow equation \eqref{eq:opflow} for $s\to\infty$. Wegner 
\cite{Wegner:1994dk} originally proposed a class of generators of the form
\begin{equation}\label{eq:def_Wegner_general}
  \etaO(s) \equiv \comm{H_d(s)}{H_{od}(s)}\,,
\end{equation}
that is widely used in applications, although it gives rise to stiff flow equations,
and more efficient alternatives exist for specific applications \cite{Bogner:2010pq,Hergert:2016jk,Hergert:2017kx}.
Wegner generators are constructed by splitting the Hamiltonian into suitably chosen 
\emph{diagonal} ($H_d(s)$) and \emph{off-diagonal} ($H_{od}(s)$) parts. These labels 
are a legacy of applying this generator to drive finite-dimensional matrices towards 
diagonality. For our purposes, they reflect the desired structure of the operator in 
the limit $s\to\infty$: We want to keep the diagonal part and drive $H_{od}(s)$
to zero by evolving it via Eq.~\eqref{eq:opflow} (see Refs.~\cite{Wegner:1994dk,Kehrein:2006kx,Bogner:2010pq,Hergert:2016jk,Hergert:2017kx}). 

To implement the operator flow equation \eqref{eq:def_Hod}, we need to express
$\eta(s)$ and $H(s)$ in a basis of suitable operators $\{O_i\}_{i\in\NS}$, 
\begin{align}
  \eta(s) &= \sum_{i}\eta_i(s) O_i\,,\\
  H(s) &= \sum_i H_i(s) O_i(s)\,,\label{eq:H_couplings}
\end{align}
where $\eta_i(s)$ and $H_i(s)$ are the running couplings of the operators. If
the algebra of the operators $O_i$ is closed naturally or with some truncation,
we have
\begin{align}
  \comm{O_i}{O_j} = \sum_kc_{ijk}O_k \;\; (+ \ldots)
\end{align}
and Eq.~\eqref{eq:opflow} becomes a system of flow equations for the coupling
coefficients: 
\begin{equation}\label{eq:flow_couplings}
  \totd{}{s}H_i(s) = f_i(\boldsymbol{c},\boldsymbol{\eta}(s),\boldsymbol{H}(s))\,,
\end{equation}
where the bold quantities collect the algebra's structure constants 
and the running couplings, respectively. From this discussion, it is clear that the choice 
of the $O_i$ can have a significant effect on the size of the system of flow 
equations, as well as the quality of any introduced truncations.

An important application of the SRG in nuclear many-body theory is the
dialing of the operators' resolution scales. This is achieved by using the
Wegner-type generator 
\begin{equation}\label{eq:mom_generator}
  \eta(\lambdaSRG)=\comm{T}{H(\lambdaSRG)}
\end{equation}
to band-diagonalize the Hamiltonian in momentum space, and thereby decouple
low- and high-momentum physics in the operators and eigenstates. As indicated in Eq.~\eqref{eq:mom_generator} the flow is typically re-parameterized by $\lambdaSRG=s^{-1/4}$, which characterizes the width of 
the band in momentum space and controls the magnitude of the momentum 
transferred in an interaction process. For example, $|\vec{k}_i - \vec{k}_f|\lesssim \lambdaSRG$ 
in a two-nucleon system  \cite{Bogner:2010pq,Hergert:2017bc}. 

Nowadays, the momentum space evolution is regularly performed for two- and three-nucleon 
forces \cite{Jurgenson:2009bs,Bogner:2010pq,Hebeler:2012ly,Wendt:2013ys,Calci:2014xy}.
In light of the previous discussion, it can be understood as choosing the operator 
basis
\begin{equation}\label{eq:free_space_basis}
  \mathcal{B} = \{\aaO_p\aO_q,\; \aaO_p\aaO_q\aO_s\aO_r,\; \aaO_p\aaO_q\aaO_r\aO_u\aO_t\aO_s,\;\ldots\}_{pqrstu\ldots\in \NS}\,,
\end{equation}
with creation and annihilation operators referring to (discretized) single-particle
momentum modes, and truncating four- and higher-body terms that appear when the
commutators of the basis operators are evaluated. Since the commutator of an $M$-body
and an $N$-body operator in the basis \eqref{eq:free_space_basis} acts at least 
on $K=\max(M,N)$ particles, the SRG evolution is exact for $A\leq 3$ systems under
this truncation \cite{Jurgenson:2009bs,Wendt:2013ys}. It is implemented by working
with the matrix representations of $H(s)$ in two- and three-nucleon systems,
whose entries correspond to the coupling constants in our chosen operator basis 
(cf.~Eq.~\eqref{eq:H_couplings}). For efficiency, an additional basis change is 
made to center-of-mass and relative coordinates.

In principle, the strategy for evolving nuclear interactions towards some form of
``diagonality'' could be used to determine eigenvalues of many-body Hamiltonians,
but the computational cost for dealing either with exponentially growing matrix 
representations or induced terms of high particle rank is prohibitive. This motivates 
the implementation of the flow equation with a different choice of basis operators
in the In-Medium SRG (see Section \ref{sec:imsrg}).

\subsection{Many-Body Methods}
\label{sec:many_body}

Let us now discuss commonly used many-body methods for solving the nuclear Schr\"{o}dinger
equation. Roughly speaking, they fall into two categories: configuration space 
methods that expand the nuclear eigenstates on a basis of known many-body states, 
or coordinate-space methods that work directly with the wave function and optimize 
them in some fashion. Our goal is to use approaches that systematically converge 
to an exact result, e.g., by adding more and more particle-hole excitations of a 
selected reference state to the many-body basis of a configuration space, or by 
exhausting the distribution of meaningful wave function parameters.

The discussion in the following sections will be light on mathematical details, which 
can be found in more specialized articles and reviews, including other contributions
to the present volume. The goal is to review only certain ideas that will become 
relevant later on.

\subsubsection{The Many-Body Problem in Configuration Space}
\label{sec:config_space}
Let us briefly discuss the general setup of the configuration-space approaches.
We choose a single-particle basis, e.g., the eigenstates of a harmonic oscillator, 
and use it to construct a basis of Slater determinants for the many-body Hilbert
space. Usually, the many-body basis is organized by selecting a reference
state $\ket{\Phi}$ and constructing its particle-hole excitations in order to 
account for the natural energy scales of the system under consideration. For 
further use, we define
\begin{equation}
  \ket{\Phi^{a\ldots}_{i\ldots}}\equiv\{\aaO_a\ldots\aO_i\ldots\}\ket{\Phi}\,,
\end{equation} 
where particle ($a,b,\ldots$) and hole ($i,j,\ldots$) indices run over unoccupied 
and occupied single-particle states, respectively\footnote{This labeling scheme is 
commonly used in chemistry \cite{Shavitt:2009}, and it is used with increasing frequency
in nuclear physics as well.}. The parentheses indicate 
that the strings of creation and annihilation operators are \emph{normal ordered} 
with respect to the reference state. They are related to the original operators by
\begin{align}
  \aaO_p\aO_q &= \{\aaO_p \aO_q\} + C_{qp}\,,\label{eq:no1b}\\
  \aaO_p\aaO_q\aO_s\aO_r &= \{\aaO_p \aaO_q\aO_s \aO_r\} 
    + C_{rp}\{\aaO_q\aO_s\} 
    - C_{sp}\{\aaO_q \aO_r\}
    + C_{sq}\{\aaO_p \aO_r\}
    - C_{rq}\{\aaO_p \aO_s\}\notag\\
    &\hphantom{=}+ C_{rp}C_{sq} - C_{sp}C_{rq}\,,\label{eq:no2b}
\end{align}
where the indices $p,q,\ldots$ run over all single-particle states, and 
the \emph{contractions} are defined as
\begin{equation}
    C_{qp} \equiv \matrixe{\Phi}{\aaO_p\aO_q}{\Phi} = \rho_{qp}
\end{equation}
(see, e.g., Refs.~\cite{Hergert:2016jk,Hergert:2017kx} 
for more details).

\begin{figure}[t]
  \setlength{\unitlength}{\textwidth}
  \begin{center}
    \includegraphics[width=0.75\unitlength]{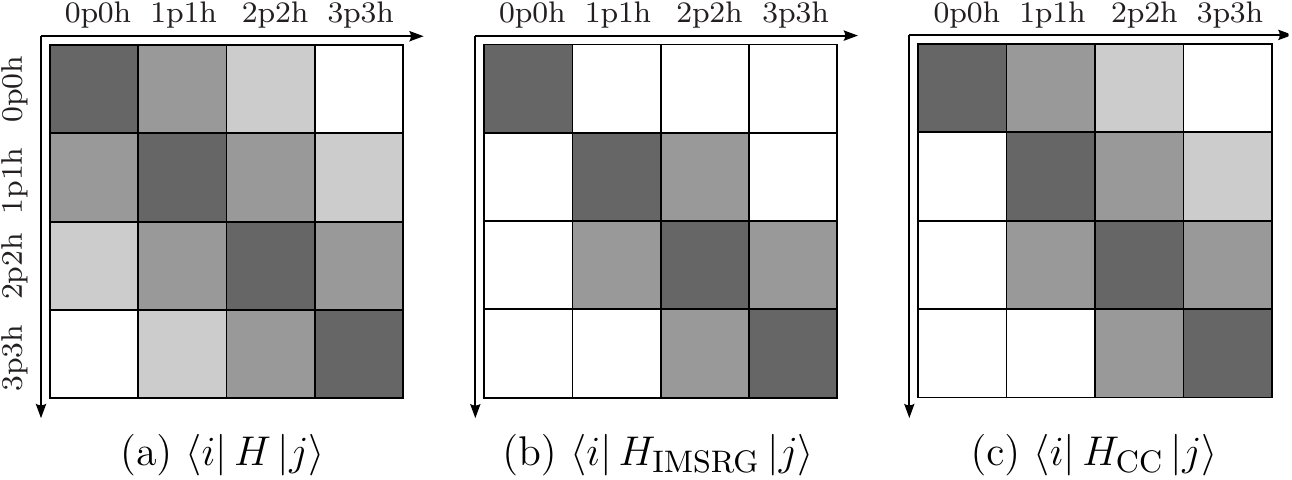}
  \end{center}
  \caption{
  \label{fig:imsrg_cc}
  Decoupling of particle-hole excitations from a 0p0h reference state:  the schematic matrix representation of the initial Hamiltonian $H_0$ (a) and the transformed Hamiltonians obtained from IMSRG  (b) and CC (c), respectively. (See text for details.)}
\end{figure}

Let us now consider a Hamiltonian containing up to two-body interactions, for simplicity.
In normal-ordered form, it is given by
\begin{equation}\label{eq:def_H}
  \HO = E_0 + \sum_{pq}f_{pq} \{\aaO_p\aO_q\} + \frac{1}{4}\sum_{pqrs}\Gamma_{pqrs} \{\aaO_p\aaO_q\aO_s\aO_r\}\,,
\end{equation}
where $E$ is the energy expectation value of the reference state, while $f$
and $\Gamma$ are the mean-field Hamiltonian and residual two-body interaction, 
respectively \cite{Hergert:2016jk,Hergert:2017kx}. Our
task is to solve the many-body Schr\"{o}dinger equation for this Hamiltonian
to determine its eigenvalues and eigenstates, either in an approximate fashion
or by exactly diagonalizing its matrix representation, which is shown in 
Fig.~\ref{fig:imsrg_cc}(a).

\subsubsection{Many-Body Perturbation Theory}
\label{sec:mbpt}

Many-Body Perturbation Theory (MBPT) is the simplest configuration-space approach for 
capturing correlations in interacting quantum many-body systems. It has enjoyed widespread 
popularity in treatments of the many-electron system since the early days of quantum 
mechanics, and it comes in a myriad of flavors (see, e.g., Ref.~\cite{Kutzelnigg:2009ly}
and references therein). 
A major factor in its success is that the Coulomb interaction is sufficiently weak to
make perturbative treatments feasible. Applications in nuclear physics had long been
hindered by the strong short-range repulsion and tensor interactions in realistic nuclear
forces, despite the introduction of techniques like Brueckner's $G$ matrix formalism
that were meant to resum the strong correlations from these contributions 
\cite{Brueckner:1954qf,Brueckner:1955rw,Day:1967zl,Brandow:1967tg}. These issues were overcome with 
the introduction of the SRG evolution to low resolution scales, which makes nuclear 
interactions genuinely perturbative, albeit at the cost of inducing three-and higher 
many-body interactions \cite{Bogner:2010pq}. As a consequence, MBPT has undergone a 
renaissance in nuclear physics in the past decade \cite{Tichai:2020ft}, leading to 
efficient applications for the computation of ground-state properties 
\cite{Roth:2010ys,Langhammer:2012uq,Tichai:2016vl} and the
construction of effective Shell Model interactions and operators (see, e.g., 
Refs.~\cite{Otsuka:2010cr,Holt:2013ec,Tsunoda:2014fk,Holt:2014vn}, or the reviews 
\cite{Coraggio:2009hb,Stroberg:2019th} and references therein). These successes have
also motivated the development of novel types of MBPTs 
\cite{Tichai:2018kx,Tichai:2018bq,Tichai:2020ft}.

In a nutshell, MBPT assumes that the Hamiltonian can be partitioned into a
solvable part $\HO_0$ and a perturbation $\HO_I$,
\begin{equation}
  \HO = \HO_0 + \HO_I\,,
\end{equation}
which then allows an order-by-order expansion of its eigenvalues and eigenstates
in powers of $\HO_I$, usually starting from a mean-field solution. In the 
Rayleigh-Schr\"{o}dinger formulation of MBPT, which is widely used for its convenience, 
\begin{align}
  \ket{\Psi} &= \ket{\Phi} + \sum_{n=1}^\infty\left(\frac{\HO_I}{\HO_0 - E^{(0)}}\right)^n\ket{\Phi}\,,\label{eq:mbpt_state}\\
  E &= E^{(0)} +  \sum_{n=0}^\infty\bra{\Phi}\HO_I\left(\frac{\HO_I}{\HO_0 - E^{(0)}}\right)^n\ket{\Phi}\,,\label{eq:mbpt_energy}
\end{align}
where $E^{(0)}$ is the unperturbed energy. If we assume that the reference Slater
determinant $\ket{\Phi}$ has been variationally optimized by solving the Hartree-Fock
equations, $E_0$ in Eq.~\eqref{eq:def_H} is the Hartree-Fock energy and $f$ is diagonal.
Then we can introduce the so-called M\o{}ller-Plesset partitioning,
\begin{equation}
  \HO_0 = E_0 + \sum_{p}f_{p}\{\aaO_p\aO_p\} \,,\quad \HO_I = \frac{1}{4}\sum_{pqrs}\Gamma_{pqrs} \{\aaO_p\aaO_q\aO_s\aO_r\}\,,
\end{equation}
and note that the Slater determinants of the basis introduced in Section \ref{sec:config_space}
are eigenstates of $\HO_0$:
\begin{equation}
  \HO_0 \ket{\Phi^{a\ldots}_{i\ldots}} = (E_0 + f_a + \ldots - f_i - \ldots )\ket{\Phi^{a\ldots}_{i\ldots}}\,.
\end{equation}
The eigenvalues of $\HO_0$ then become the unperturbed energies appearing in Eqs.~\eqref{eq:mbpt_state},\eqref{eq:mbpt_energy},
and the energy including a finite number of correction terms can be evaluated straightforwardly. 
For example, the ground-state energy through second order is given by
\begin{equation}\label{eq:mbpt2}
  E = E_0 - \frac{1}{4}\sum_{abij}\frac{|\Gamma_{abij}|^2}{f_a + f_b - f_i - f_j }\,.
\end{equation}
For a more detailed discussion, we refer to Ref.~\cite{Tichai:2020ft} and references therein.

The expression \eqref{eq:mbpt2} can serve to illustrate both advantages and drawbacks of
an MBPT treatment of nuclei. We see that the second-order energy can be evaluated very
efficiently, since it requires a non-iterative calculation whose computational effort
scales polynomially in the single-particle basis size $N$, namely as $\mathcal{O}(N^4)$. 
The reason is that the construction of the Hamiltonian matrix (Fig.~\ref{fig:imsrg_cc}(a)) can 
be avoided. In fact, the computational scaling is even more favorable, because we can distinguish 
particle and hole states and achieve $\mathcal{O}(N_p^2N_h^2)$, and we typically have 
$N_h \sim A \ll N_p$. Although there is a proliferation of terms with increasing order
\cite{Kucharski:1986fk,Shavitt:2009,Tichai:2020ft}, MBPT is still fundamentally 
polynomial and therefore more efficient than an exact diagonalization, whose cost scales 
exponentially with $N$. It is also clear from Eq.~\eqref{eq:mbpt2} that the expansion
of the exact eigenvalue will break down if one (or more) of the energy denominators 
become small due to (near-)degeneracies of the unperturbed energies. Thus, MBPT works
best for ground states in systems with a strong energy gap, i.e., closed-shell nuclei, 
although extensions for more complex scenarios exist (see Refs.~\cite{Brandow:1967tg,Shavitt:2009,Tichai:2020ft} 
and references therein). A noteworthy new development is Bogoliubov MBPT, in which 
particle number symmetry is broken and eventually restored 
\cite{Tichai:2018kx,Ripoche:2020cz,Demol:2020nm}.

As mentioned at the beginning of this section, MBPT can be used to derive effective 
interactions and operators. The primary tool for such efforts is the $\hat{Q}$-box 
or folded-diagram resummation of the perturbative series 
(see Refs.~\cite{Hjorth-Jensen:1995ys,Coraggio:2009hb,Stroberg:2019th} and references therein).

\subsubsection{In-Medium Similarity Renormalization Group}
\label{sec:imsrg}

As already mentioned in our discussion of the SRG in Section \ref{sec:srg}, we could
envision applying SRG techniques not only to preprocess the nuclear interactions,
but also to compute eigenvalues and eigenstates. For all but the lightest nuclei, 
applying the SRG to the Hamiltonian matrix is hopeless, so we work with the 
operators instead.

Let us again consider the matrix representation shown in Fig.~\ref{fig:imsrg_cc}~(a).
We want to design a transformation that will decouple the one-dimensional 0p0h block 
in the Hamiltonian matrix, spanned by a reference state Slater determinant $\ket{\Phi}$, 
from all excitations as the flow equation \eqref{eq:opflow} is integrated. The matrix element 
in this block will then be driven towards an eigenvalue (up to truncation errors), 
and the unitary transformation becomes a mapping between the reference Slater determinant 
and the exact eigenstate (see below). In principle, we could use a suitably chosen 
reference to target different eigenstates, e.g., by taking references which are 
expected to have a large overlap with the target state (see Section 10.3 in Ref.~
\cite{Hergert:2017bc}). In practice, we usually target the ground state 
by using a Hartree-Fock Slater determinant as our reference.

To implement the operator flow, we need to choose an operator basis to express
$\HO(s)$ and the generator $\etaO(s)$. Instead of using the basis \eqref{eq:free_space_basis}, 
we switch to operators that are normal ordered with respect to the reference state 
$\ket{\Phi}$:
\begin{equation}\label{eq:imsrg_basis}
  \mathcal{B} = \left\{ \{\aaO_p\aO_q\}, \{\aaO_p\aaO_q\aO_s\aO_r\}, \{\aaO_p\aaO_q\aaO_r\aO_u\aO_t\aO_s\}, \ldots\right\}_{pqrstu\ldots\in\NS}\,.
\end{equation}
Commutators of these operators can feed into terms of \emph{lower} particle rank: 
For instance, a commutator of $M$-body and $N-$body operators generates 
$|M-N|$-body through $(M+N-1)$-body operators, while the lower bound for the basis
\eqref{eq:free_space_basis} is $\max(M,N)$ (cf.~Section \ref{sec:srg}). As a result, 
the complexity of the flow equations for the operators' coupling coefficients  
increases due to the appearance of additional terms that depend on the contractions 
introduced in Eqs.~\eqref{eq:no1b} and \eqref{eq:no2b}. These contractions translate 
into density matrices (or occupation numbers) --- hence the name In-Medium SRG. At the 
same time, we achieve a reduction of the truncation error because only the 
\emph{residual}, contraction-independent parts of the operators \eqref{eq:no1b} and 
\eqref{eq:no2b} are omitted. In the majority of applications to date, we truncate 
all operators and their commutators at the two-body level, defining the IMSRG(2) 
truncation scheme. More details can be found in 
Refs.~\cite{Hergert:2016jk,Hergert:2017kx,Hergert:2017bc,Stroberg:2019th}.

In the chosen basis we now identify the parts of the Hamiltonian that are 
responsible for coupling the reference state to 1p1h and 2p2h excitations, and define 
the off-diagonal Hamiltonian (cf.~\ref{sec:srg}) as 
\begin{equation}\label{eq:def_Hod}
  \HO_{od} \equiv \sum_{ai} f_{ai} \{\aaO_a\aO_i\} + \frac{1}{4}\sum_{abij} \Gamma_{abij} \{\aaO_a\aaO_{b}\aO_{j}\aO_{i}\} + \text{H. c.}\,.
\end{equation}
We use this $\HO_{od}$ to construct a generator, either using Wegner's ansatz 
\eqref{eq:def_Wegner_general} or an alternative choice \cite{Hergert:2016jk,Hergert:2017kx}. 
Plugging the generator into the operator flow equation \eqref{eq:opflow}, we obtain
a system of flow equations for the energy $E(s)$ and the coefficients $f_{pq}(s), 
\Gamma_{pqrs}(s), \ldots$ (cf.~Eq.~\eqref{eq:flow_couplings} and Refs.~\cite{Hergert:2016jk,Hergert:2017kx,Stroberg:2019th}). 
By integrating these flow equations, we evolve the Hamiltonian operator so that 
its matrix representation assumes the shape shown in Fig.~\ref{fig:imsrg_cc}~(b). 
We note that the suppression of $\HO_{od}$ not only leads to the desired ground-state 
decoupling, but also eliminates the outermost band in the Hamiltonian matrix. 
This simplification makes the evolved Hamiltonian an attractive input for other 
approaches, e.g., configuration interaction (CI) or equation-of-motion methods 
(see Refs.~\cite{Parzuchowski:2017yq,Stroberg:2017sf,Stroberg:2019th,Gebrerufael:2017fk,Yao:2018wq,Hergert:2018th,Yao:2020mw}
and discussion below).

\noindent\textbf{Valence-Space IMSRG.} Soon after introducing the IMSRG in 
nuclear physics \cite{Tsukiyama:2011uq}, 
Tsukiyama, Bogner and Schwenk proposed the use of the IMSRG flow to derive 
Hamiltonians (and other effective operators) for use in nuclear Shell Model
calculations \cite{Tsukiyama:2012fk}. This is achieved by partitioning 
the single-particle basis into core, valence, and beyond-valence states, 
normal ordering all operators with respect to a Slater determinant describing
the closed-shell core, and extending the definition of the off-diagonal 
Hamiltonian \eqref{eq:def_Hod} to include all terms that couple valence and 
non-valence states. The eigenvalue problem for the evolved Hamiltonian can
then be solved in the valence space with widely available Shell model codes
\cite{Caurier:2005qf,Brown:2014fk,Engeland:2017oa,Johnson:2018hrx,Shimizu:2019om}.
After a study of the oxygen isotopic chain revealed an increasing overbinding 
away from the chosen core \cite{Bogner:2014tg}, we adopted a normal-ordering 
scheme that uses an ensemble of Slater determinants to account for partially 
filled shells in open-shell nuclei \cite{Stroberg:2016fk,Stroberg:2017sf}. 
This improved operator basis, along with the valence decoupling procedure
and subsequent Shell Model diagonalization defines what is nowadays called 
the valence-space IMSRG (VS-IMSRG) --- see Ref.~\cite{Stroberg:2019th} for a 
recent review.

\noindent\textbf{Correlated Reference States and Multi-Reference IMSRG.}
Another important development was the extension of the IMSRG formalism
to correlated reference states, in the so-called Multi-Reference IMSRG
(MR-IMSRG) \cite{Hergert:2013ij,Hergert:2017kx,Hergert:2017bc}. The unitarity 
of the IMSRG transformation allows us to control to what extent correlations 
are described by either 
the Hamiltonian or the reference state. We can see this by considering the
stationary Schr\"{o}dinger equation and applying $\UO(s)$:
\begin{equation}
  \left[\UO(s)\HO\UUO(s)\right] \UO(s)\ket{\Psi_k} = E_k \UO(s) \ket{\Psi_k} \,.
\end{equation}
The transformation shifts correlations from the wave function into the evolved, 
RG-improved Hamiltonian $\HO(s)=\UO(s)\HO\UUO(s)$, and any many-body method that
uses this Hamiltonian as input now needs to describe $\UO(s)\ket{\Psi_k}$, which
should be less correlated than the exact eigenstate $\ket{\Psi_k}$. In the
extreme cases, $\UO(s)=\idO$ and the wave function carries all correlations,
or $\UO(s)$ has shifted all correlations into the Hamiltonian and 
$\ket{\Phi}=\UO(s)\ket{\Psi}$ is a simple Slater determinant.

\begin{figure}[t]
  \setlength{\unitlength}{\textwidth}
  \begin{center}
    \includegraphics[width=0.5\unitlength]{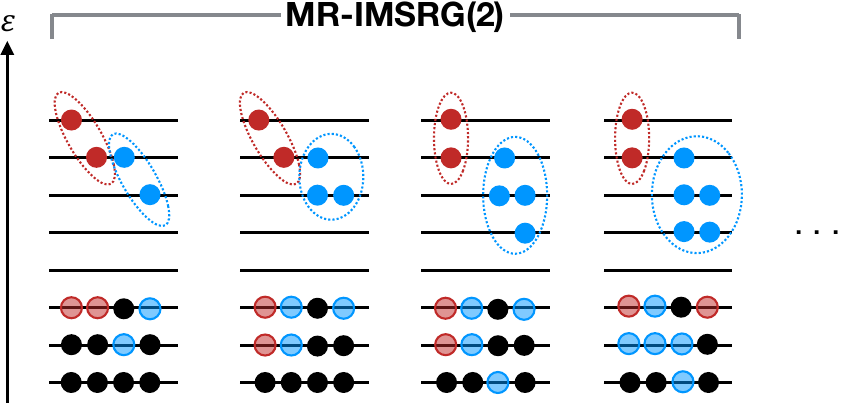}
  \end{center}
  \caption{\label{fig:static_corr}
    Schematic view of correlations in nuclei. Solid
    circles indicate nucleons, transparent circles hole states, and dashed
    ellipses indicate correlations between nucleons. Certain 2p2h, 3p3h and 
    higher correlations (indicated in blue) are built into a correlated
    wave function that then serves as the reference state for an MR-IMSRG(2)
    calculation (capturing correlations indicated in red), while up to 
    an IMSRG(A) calculation would be needed for an equivalent description
    in the conventional framework.
  }
\end{figure}

Correlated reference states can be particularly useful for the description of
systems with strong \emph{static} or \emph{collective} correlations, like 
open-shell nuclei with strong intrinsic deformation or shape coexistence. 
Reference states that describe these types of correlations efficiently, e.g.,
through symmetry breaking and restoration (also see Section \ref{sec:cc}), 
are an ideal complement to the IMSRG transformation, which excels at capturing 
\emph{dynamic} correlations, involving the excitation of a few particles up to 
high energies. This complementarity is schematically illustrated in 
Fig.~\ref{fig:static_corr}: Collective correlations that would require as 
much as an IMSRG(A) calculation in the conventional approach are built into
the reference state, and an MR-IMSRG(2) calculation is sufficient to treat
the bulk of the dynamical correlations in the system.

Reference state correlations are built into the MR-IMSRG framework by 
using a generalized normal ordering \cite{Kutzelnigg:1997fk,Kong:2010kx,Hergert:2017kx} 
that is extended with contractions of higher rank, namely the \emph{irreducible $k$-body 
density matrices $\lambda^{(k)}$}:
\begin{align}
  \lambda_{pq} &\equiv \rho_{pq} \,,\\
  \lambda_{pqrs} &\equiv \rho_{pqrs} - \rho_{pr}\rho_{qs} + \rho_{qr}\rho_{ps}\,,\label{eq:def_Lambda2}
\end{align}
etc. The irreducible densities matrices encode the correlation content of 
an arbitrary reference state $\ket{\Phi}$, hence they vanish for Slater 
determinants. While the basis of normal-ordered operators superficially
is the same as in the conventional IMSRG, shown in Eq.~\eqref{eq:imsrg_basis},
the inclusion of the irreducible densities (cf.~Eqs.~\eqref{eq:no1b} and \eqref{eq:no2b})
equips the basis with the capability to describe the correlations that
are present in the reference state, which in turn should help to reduce
MR-IMSRG truncation errors. To understand this, let us assume that we know
the ground state of our system, and we normal order the Hamiltonian with 
respect to this correlated state. Then the zero-body part of the normal
ordered Hamiltonian already is the exact ground-state energy, and the 
normal-ordered one-, two- and higher-body parts do not matter at all for 
our result, and neither does their evolution under an exact or truncated
MR-IMSRG flow. Thus, the better the reference state matches the ground 
state, the less work the MR-IMSRG evolution and any subsequent many-body 
method have to do to obtain the correct ground-state energy.

\noindent\textbf{Computational Scaling and Magnus Expansion.}
The computational scaling of all three IMSRG flavors discussed here ---
traditional, VS-IMSRG, and MR-IMSRG --- is governed by the truncation
scheme. If we truncate operators and commutators at the two-body level,
as briefly mentioned above, the number of flow equations scales as 
$\mathcal{O}(N^4)$ with the single-particle basis size $N$, and the 
computational effort for evaluating the right-hand sides as $\mathcal{O}(N^6)$.
This holds despite the greater complexity of the MR-IMSRG flow 
equations, which contain terms containing irreducible two- and 
higher-body density matrices.

Any observables of interest must, in principle, be evolved alongside
the Hamiltonian for consistency, which would create a significant overhead.
In practice, we can address this issue by using the so-called Magnus 
formulation of the IMSRG \cite{Morris:2015ve,Parzuchowski:2017yq,Hergert:2017bc,Stroberg:2019th}:
Assuming that the IMSRG transformation can be written as an explicit exponential,
$U(s)=\exp\Omega(s)$, we can solve a single set of flow equations for the
anti-Hermitian operator $\Omega(s)$ instead of evolving observables 
separately. All operators of interest can then be computed by applying
the Baker-Campbell-Hausdorff expansion to $O(s)=\exp[\Omega(s)] O\exp [-\Omega(s)]$.

\noindent\textbf{IMSRG Hybrid Methods.}
As noted earlier in this section, the conventional IMSRG evolution 
makes the matrix representation of the Hamiltonian more diagonal
by suppressing couplings between the npnh excitations of the reference 
state. This implies a decoupling of energy scales of the many-body
system, analogous to the decoupling of momentum scales by the free-space 
SRG, although there are differences in detail that are associated with
the operator bases in which the flow is expressed (cf.~Eqs.~\eqref{eq:free_space_basis},
and \eqref{eq:imsrg_basis}). 

From this realization, it is not a big step to consider using the IMSRG 
to construct RG-improved Hamiltonians for applications in other methods,
defining novel hybrid approaches. In fact, even the original IMSRG 
formulation can be understood from this perspective: The evolution generates 
a Hamiltonian that yields the exact ground-state energy (up to truncations) 
in a Hartree-Fock calculation, except the HF equations are automatically 
satisfied for the evolved $\HO$, and we can read off the ground-state energy 
directly. The same Hamiltonian can then be used as input for EOM methods 
to compute excitation spectra \cite{Parzuchowski:2017yq}. Likewise, the 
VS-IMSRG produces an RG-improved Hamiltonian that serves as input for a 
Shell Model diagonalization.

Applying the same logic as in the VS-IMSRG case, the IMSRG has been merged
with the No-Core Shell Model (NCSM, see Section \ref{sec:ncsm}) into the
In-Medium NCSM \cite{Gebrerufael:2017fk,DAlessio:2020mg}. In this approach,
the IMSRG improves the Hamiltonian with dynamical correlations from high-energy 
few-nucleon excitations that would require enormously large model spaces
in the conventional NCSM, and the exact diagonalization in a 
small model space describes the dynamics of many-nucleon excitations. 
The NCSM as the ``host'' method is rooted in the same particle-hole expansion
picture as the IMSRG itself, but this is not a requirement. Another new hybrid 
method is the In-Medium Generator Coordinate Method (IM-GCM), which relies on 
the GCM as a host method to capture collective correlations \cite{Yao:2018wq,Hergert:2018th,Yao:2020mw}. 
In this approach, a many-body basis is generated by restoring the symmetries 
of mean field solutions with various types of shape and gauge configuration
constraints, which is very different from the particle-hole excitation basis 
discussed so far.

\subsubsection{Coupled Cluster Methods}
\label{sec:cc}

The Coupled Cluster (CC) method \cite{Shavitt:2009,Hagen:2014ve} is an
older cousin of the IMSRG approach. It can also be understood as a decoupling
transformation of the Hamiltonian, but in contrast to the IMSRG, it relies on
a non-unitary similarity transformation (see Fig.~\ref{fig:imsrg_cc}). Traditionally, 
CC is motivated by an exponential ansatz for the exact wave function of a system,
\begin{equation}\label{eq:cc_wf}
  \ket{\Psi_{CC}}= e^T\ket{\Phi}\,,
\end{equation}
where $\ket{\Phi}$ is a reference Slater determinant, and $T$ is the so-called 
\emph{cluster operator}. This operator is expanded on particle-hole excitations,
\begin{equation}\label{eq:def_tcc}
  T=\sum_{ph} t_{ai}\{\aaO_a\aO_i\} + \frac{1}{4}\sum_{abij}t_{abij}
  \{\aaO_a\aaO_{b}\aO_{j}\aO_i\} + \ldots\,,
\end{equation}
with the \emph{cluster amplitudes} $t_{ai}, t_{abij}, \ldots$. In practical applications, 
the $T$ is truncated to include up to 2p2h (CC with Singles and Doubles, or CCSD) or 
3p3h terms (CCSDT, including Triples). Various schemes exist for iteratively or non-iteratively
including subsets of Triples \cite{Taube:2008kx,Taube:2008vn,Shavitt:2009,Binder:2013fk,Hagen:2014ve}.
When it acts on the reference state $\ket{\Phi}$, $e^T$ admixes arbitrary powers of 
few-particle, few-hole excitations. Note, however, that the cluster operator $T$ is 
not anti-Hermitian because it lacks de-excitation operators, and therefore $e^T$ is 
not unitary. 

The cluster amplitudes are determined by demanding that the transformed Hamiltonian,
\begin{equation}\label{eq:def_HCC}
  H_{CC} \equiv e^{-T}He^T\,,
\end{equation}
does not couple the reference to 1p1h and 2p2h states (see Fig.~\ref{fig:imsrg_cc}). 
Using notation introduced in Section \ref{sec:config_space}, the decoupling conditions 
lead to the following system of non-linear equations:
\begin{align}\label{eq:cc_eqs_E}
  \bra{\Phi}e^{-T}He^T\ket{\Phi}&=E_{CC}\,,\\
  \bra{\Phi^{a}_{i}}{e^{-T}He^T}\ket{\Phi} &=0\,,\label{eq:cc_eqs_1p1h}\\
  \bra{\Phi^{ab}_{ij}}{e^{-T}He^T}\ket{\Phi} &=0 \label{eq:cc_eqs_2p2h}\,.
\end{align}
Here, $E_{CC}$ is the CC ground-state energy, which corresponds to the one-dimensional
block in the upper left of Fig.~\ref{fig:imsrg_cc}~(c) and is analogous to the zero-body
part of the IMSRG-evolved Hamiltonian, as discussed in the previous section. The other
blocks in the first column of the matrix vanish because of the CC equations 
\eqref{eq:cc_eqs_E}--\eqref{eq:cc_eqs_2p2h}.

Since the CC transformation is non-unitary, one needs to be careful when
one evaluates observables using the CC wave function, or uses $H_{CC}$ as 
input for equation-of-motion calculations or other applications
 \cite{Shavitt:2009,Hagen:2014ve}. For instance,
the non-Hermiticity of $H_{CC}$ forces us to consider left and right
eigenstates separately. This is a drawback compared to unitary transformation 
methods like the IMSRG. Coupled Cluster also has advantages, though: For
instance, the Baker-Campbell-Hausdorff expansion appearing in 
Eqs.~\eqref{eq:cc_eqs_E}--\eqref{eq:cc_eqs_2p2h} automatically terminates at 
finite order because the cluster operator only contains excitation operators.
For the same reason, Eq.~\eqref{eq:cc_eqs_1p1h} will automatically solve the 
Hartree-Fock equations, so any Slater determinant is equally well suited as a
reference state, while MBPT, IMSRG and even exact diagonalization approaches
exhibit (some) reference-state dependence.

\noindent\textbf{Symmetry Breaking and Collective Correlations.} While most
applications of CC theory in nuclear physics have enforced and exploited 
spherical symmetry, the capabilities for performing $M$-scheme calculations
that allow nuclei to develop intrinsic deformation have existed for more
than a decade. This is a more natural approach for capturing collective 
correlations than the construction of Triples, Quadruples (4p4h) and ever
higher particle-hole excitations of a spherical reference (cf.~Section \ref{sec:imsrg}).
Converging such calculations is challenging because the single-particle 
basis typically grows by an order of magnitude or more, and the broken 
symmetries must eventually be restored. The formalism for symmetry restoration 
in CC has been developed in Refs.~\cite{Duguet:2015ye,Signoracci:2015mz,Duguet:2017sh,Qiu:2019rf}.
In fact, the work of Duguet et al. forms the basis of recent works on
symmetry breaking and restoration in MBPT \cite{Tichai:2018kx,Ripoche:2020cz,Demol:2020nm}.
Applications are currently underway.

\noindent\textbf{Shell-Model CC.} Like the IMSRG, the CC framework can be used to construct
effective interactions and operators for Shell model calculations. Initial
work in that direction applied Hilbert space projection techniques 
(cf.~Section \ref{sec:ncsm}) to construct a so-called CC effective interaction
(CCEI) \cite{Jansen:2014qf,Jansen:2016kq}, but the construction of the model
spaces via Equation-of-Motion CC methods proved to be computationally 
expensive. The CCEI approach is now superseded by the Shell Model CC
method \cite{Sun:2018ul}, which applies a second similarity transformation
to $H_{CC}$ in Fock space, similar to VS-IMSRG decoupling (cf.~Section \ref{sec:imsrg}).

\noindent\textbf{Unitary CC.}
While almost all applications of CC in nuclear physics use the traditional
ansatz \eqref{eq:cc_wf}, unitary CC (UCC) approaches that parameterize the
wave function as $\ket{\Psi_{UCC}} = e^{T - T^\dagger}\ket{\Phi}$ have been 
used in numerous studies in quantum chemistry (see, e.g., \cite{Taube:2006kl,Bartlett:2007kx}). 
Unitary CC wave functions have also become a popular ansatz for the Variational
Quantum Eigensolver (VQE) algorithm on current and near-term quantum devices 
\cite{Dumitrescu:2018az,Lu:2019bs}.
It is also worth noting that the recently revived Unitary Model Operator Approach 
(UMOA) is closely related to UCC \cite{Miyagi:2017hl,Miyagi:2019xv}. 

\subsubsection{Self-Consistent Green's Functions}
\label{sec:scfg}

Self-Consistent Green's Function (SCGF) theory is another prominent approach for 
solving the nuclear many-body problem with systematic approximations 
\cite{Dickhoff:2004fk,Barbieri:2017sh,Soma:2020dw,Rios:2020mj}. The Green's
Functions in question are correlation functions of the form
\begin{equation}\label{eq:def_propagator}
  g_{pq\ldots rs} \equiv \matrixe{\Psi^A_0}{\mathcal{T}[\aO_p(t_p)\aO_q(t_q)\ldots \aaO_s(t_s)\aaO_r(t_r)]}{\Psi^A_0}\,,
\end{equation}
which describe the propagation of nucleons in the exact ground state $\ket{\Psi^A_0}$
of the system. Using Wick's theorem, the exact $A$-body propagator \eqref{eq:def_propagator}
can be factorized into products of irreducible one-, two-, etc. propagators, similar
to the decomposition of density matrices briefly touched upon in Section \ref{sec:imsrg}. 
One can then formulate coupled equations of motion for propagators, and introduce truncations to 
obtain polynomially scaling methods, again somewhat analogous to IMSRG and CC. We must
remain aware that the propagators of SCGF, the induced operators of IMSRG, 
and the CC amplitudes are all different objects, and while their definitions may 
make the seem complementary to each other, there are subtle distinctions. One of these is 
that the $g^{(k)}$ are formally defined with respect to the exact wave function, 
while IMSRG and CC use definitions with respect to a reference state.

Practical implementations of the SCGF technique usually work with the Fourier
transforms of the propagators to the energy domain. One needs to solve 
integral equations of motion of the form
\begin{equation}\label{eq:gf_eom}
  g = g_0 + g_0 \Sigma g\,,
\end{equation}
where $g_0$ is the propagator of the non-interacting system and $\Sigma$ a kernel 
that encodes the particles' interactions, which is constructed using diagrammatic
techniques. For example, the one-body propagator is obtained by solving
\begin{equation}\label{eq:dyson}
  g_{pq}(\omega) = g_{pq}^{(0)}(\omega) + \sum_{rs}g^{(0)}_{pr}\Sigma_{rs}(\omega)g_{sq}\,,
\end{equation}
the so-called Dyson equation. From this propagator, one can compute the one-body
density matrix
\begin{equation}
  \rho_{pq} = \matrixe{\Psi^A_0}{\aaO_q\aO_p}{\Psi^A_0} = \int_{C^+} \frac{d\omega}{2\pi i} g_{pq}(\omega)\,,
\end{equation}
where $C^+$ indicates an integration contour in the complex upper half plane. 
Higher-body density matrices are connected to the corresponding higher-body
propagators in analogous fashion. Using the density matrices, one can then 
evaluate any operator expectation values of interest. For more details, we 
refer to the contributions \cite{Soma:2020fj,Rios:2020mj} to the present volume, 
and the works cited therein.

Current applications of SCGF techniques in nuclear physics make use of the
so-called Algebraic Diagrammatic Construction (ADC) scheme, with increasing
orders, denoted by ADC(n), converging to an exact solution. For closed-shell
nuclei, calculations up to ADC(3) are be performed regularly, which contain
correlations that are roughly comparable to IMSRG(2) with a perturbative 3p3h
correction (see Section \ref{sec:imsrg} and Refs.~\cite{Morris:2016xp,Parzuchowski:2017yq,Hergert:2018th}) 
and CCSD(T) (cf.~Section \ref{sec:cc}). Som\`{a} and collaborators have extended
the ADC scheme to open-shell nuclei by using Gor'kov Green's Functions with
explicitly broken particle number symmetry \cite{Soma:2011vn,Soma:2014fu}.
Applications of this framework have used a self-consistent second-order 
scheme, denoted Gor'kov-ADC(2), and the extension to Gor'kov-ADC(3) as
well the integration of particle-number projection to restore the broken
number symmetry are in progress \cite{Ripoche:2020cz,Soma:2020dw}.

While the computation of the Green's Functions tends to be a more involved task
than solving the IMSRG flow equations or CC amplitude equations, the propagator 
contains more information from a single computation than these other methods. 
For instance, one can immediately extract spectral information about the neighboring 
nuclei and the response of the system \cite{Raimondi:2019nt,Rocco:2018va}, which 
requires the application of additional techniques 
in the IMSRG \cite{Parzuchowski:2017yq} and CC approaches \cite{Jansen:2011tg,Jansen:2013zr,Hagen:2014ve}, 
or, indeed, the computation of the Green's Function using similarity-transformed 
operators. Furthermore, the kernels of the equations of motion \eqref{eq:gf_eom}
are energy-dependent effective interactions that govern the dynamics of 
(few-)nucleon-nucleus interactions. For example, the one-nucleon self-energy 
in Eq.~\eqref{eq:dyson} is an \emph{ab initio} version of an optical potential, 
as used in reaction theory \cite{Rotureau:2017km,Rotureau:2018ck,Idini:2019gl}.
We will return to this discussion in Section \ref{sec:dynamics}.

\subsubsection{Configuration Interaction Approaches}
\label{sec:ncsm}

\noindent\textbf{No-Core Configuration Interaction Methods.}
The most straightforward but also most computationally expensive approach to solving
the many-body Schr\"{o}dinger equation is to exactly diagonalize the Hamiltonian in a
basis of many-body states. In general, we refer to such approaches as No-Core Configuration 
Interaction (NCCI). ``No core'' makes it explicitly clear that all nucleons 
are treated as active degrees of freedom, in contrast to the nuclear Shell model
discussed below.

In light nuclei, the exact diagonalization can be directly formulated in Jacobi
coordinates, using translationally invariant harmonic oscillator \cite{Navratil:2000hf} 
or hyperspherical harmonic wave functions \cite{Barnea:2001kx,Barnea:2004cr}.
Since the construction of the basis states themselves and the matrix representation 
of the Hamiltonian becomes increasingly complicated and computationally expensive
as the particle number grows, one eventually has to switch to Slater determinants 
in the laboratory system, using a construction along the lines discussed in 
Section \ref{sec:config_space}.

A common choice for the single-particle basis in the laboratory system are spherical
harmonic oscillator (SHO) states, because they allow an exact separation of 
center-of-mass and intrinsic degrees of freedom provided one uses an energy-based
truncation for the model space \cite{Barrett:2013oq,Navratil:2016rw}. These choices
define what we specifically call the No-Core Shell Model (NCSM). A disadvantage of
using SHO orbitals is that they are not optimized to the energy scales of specific
nuclei, and they are poorly suited for describing physical features like 
extended exponential wave function tails. Other popular choices are Hartree-Fock 
single-particle states, and perturbatively \cite{Tichai:2019to} or nonperturbatively
enhanced natural orbitals \cite{Robin:2016ww,Robin:2017qv,Pillet:2017zl}. Model 
spaces built on these bases
no longer guarantee the separation of center-of-mass and intrinsic coordinates,
but fortunately, center-of-mass contaminations either remain small automatically 
\cite{Roth:2009oh}, or they can be suppressed using techniques like the Lawson method 
\cite{Gloeckner:1974gb}.

\noindent\textbf{Importance Truncation and Symmetry Adaptation.}
As indicated above, the main issue with exact diagonalization approaches is
the exponential (or greater) growth of the Hilbert space dimension, which is
proportional to $\binom{N}{A}$ with single particle basis size $N$ and particle
number $A$. A variety of strategies can be used to address this often-quoted 
``explosion'' of the basis size. One direction is to avoid the
construction of the full model space basis by applying importance-based
truncation or sampling methods, leading to the Importance-Truncated NCSM \cite{Roth:2009eu}
or Monte-Carlo (No-Core) CI approaches \cite{Otsuka:2001nx,Shimizu:2017ce}).

Another important research program is the exploration of many-body states that 
are constructed from the irreducible representations (irreps) of the symplectic 
group Sp(3,$\mathbbm{R}$), which describes an approximate emergent symmetry of finite nuclei
\cite{Launey:2016ef,Dytrych:2020db}.
An exact diagonalization in such a symmetry-adapted basis will offer a much
more efficient description of nuclear states with intrinsic deformation than 
the conventional NCSM, which would need to use massive model spaces with
many-particle-many-hole excitations. This reduction of the model space 
dimensions also allows such symmetry-adapted NCSM \cite{Launey:2016ef,Dytrych:2020db} 
and NCCI approaches \cite{Caprio:2020aa} to reach heavier nuclei than the 
conventional versions.

\noindent\textbf{Interacting Nuclear Shell Model with a Core (Valence CI).} Instead of treating all 
of the nucleons as active, one can also factorize the 
nuclear wave function by introducing an inert core and only treat the interactions 
of a smaller number of valence nucleons via appropriately transformed interactions:
\begin{equation}
  \ket{\Psi} = \ket{\Psi}_\text{core}\otimes\ket{\Psi}_\text{valence}\,.
\end{equation}
This, of course, is the traditional nuclear Shell model approach. Even with
the substantial reduction of the single-particle basis to a relatively small
number of valence orbitals, the numerical cost for an exact diagonalization
quickly becomes unfeasible for many medium-mass and heavy nuclei, especially
if one needs multi-shell valence-spaces to capture complex nuclear structure
features like coexisting intrinsic shapes.

In previous sections, we have discussed how a variety of many-body methods
can be used to derive valence-space interactions, hence it is not a surprise
that this is possible in NCCI approaches as well. One strategy is to project
solutions of no-core calculations for the core and its neighboring nuclei onto 
a valence-configuration space to extract
the effective Hamiltonian. The viability of this approach has been demonstrated
in several publications \cite{Lisetskiy:2008fk,Lisetskiy:2009uq,Dikmen:2015fk,Smirnova:2019pj},
although there are ambiguities in the extraction of the valence-space Hamiltonian, 
and the initial NCCI calculations that serve as input for the projection 
rapidly become expensive.

\noindent\textbf{Description of Continuum Effects and Nuclear Dynamics.}
An important breakthrough in \emph{ab initio} calculations for light
nuclei has been the merging of the NCSM with resonating group method
(RGM) techniques \cite{Navratil:2016rw,Kravvaris:2017yq}. This makes it
possible to
describe clustered states as well as reactions between light projectile(s) 
and targets. In the original NCSM/RGM approach, compact clusters of 
nucleons are described by NCSM states, which are then used to construct 
a basis of configurations $\ket{\chi_i}$ that place such clusters at 
different relative distances. In this basis, one can then solve the 
generalized eigenvalue problem, known as the Griffin-Hill-Wheeler 
equation \cite{Ring:1980bb} in the RGM context:
\begin{equation}\label{eq:ncsm_rgm}
  \mathcal{H}\ket{\Psi} = E \mathcal{N}\ket{\Psi}\,,
\end{equation}
where $\mathcal{H}$ and $\mathcal{N}$ are the so-called Hamiltonian
and norm kernels. The latter appears because the chosen basis configurations
are not orthogonal in general. The dimension of Eq.~\eqref{eq:ncsm_rgm}
is typically small, certainly compared to the NCSM model space, but
the computation of the kernels is computationally expensive since it
relies on the construction of up to three-body transition density 
matrices. In recent years, the NCSM/RGM has been extended to the
NCSM with Continuum (NCSMC), which accounts for the coupling between 
the NCSM and RGM sectors of the many-body basis \cite{Navratil:2016rw}. 
It requires solving the generalized eigenvalue problem 
\begin{equation}\label{eq:ncsmc}
  \begin{pmatrix}
    h &  \bar{h} \\
    \bar{h}  &  \mathcal{H}
  \end{pmatrix}
  \begin{pmatrix}
    \Phi \\ \chi
  \end{pmatrix}
  = E 
  \begin{pmatrix}
    \mathbbm{1} &  \bar{n} \\
    \bar{n}  &  \mathcal{N}
  \end{pmatrix}
  \begin{pmatrix}
    \Phi \\ \chi
  \end{pmatrix}\,,
\end{equation}
where $h$ and $\mathbbm{1}$ are the Hamiltonian and norm kernel in the 
NCSM sector (the latter being diagonal), $\mathcal{H}$ and $\mathcal{N}$
 the corresponding kernels in the RGM sector (cf.~Eq.~\eqref{eq:ncsm_rgm}),
and $\bar{h}$ and $\bar{n}$ encode the coupling between the sectors of the
basis. 

Alternative approaches to the description of continuum effects in the
NCSM are the Single-State HORSE (Harmonic Oscillator Representation of Scattering
Equations) method \cite{Shirokov:2016tw,Shirokov:2018tx,Bang:2000bx},
for which the nomen is omen, as well as the No-Core Gamow Shell Model (GSM), a 
no-core CI approach that constructs Slater determinants from a single-particle
Berggren basis \cite{Berggren:1968hb} consisting of bound, resonant and scattering 
states \cite{Michel:2009oz,Papadimitriou:2013dt,Rotureau:2013km,Shin:2017ef}.

\subsubsection{Quantum Monte Carlo}
\label{sec:qmc}

The most commonly used Quantum Monte Carlo (QMC) techniques in nuclear
physics make use of many-body wave functions in coordinate space representation
\cite{Carlson:2015lq,Lynn:2019dw,Gandolfi:2020fv,Tews:2020hl}. As such, they are well suited 
for the description of nuclear states with complex intrinsic structures, and 
they can readily use interactions with a high momentum cutoff, as opposed to
the configuration space methods which would exhibit poor convergence in such
cases. This allows QMC calculations to explore physics across the interfaces
of the hierarchy of EFTs for the strong interaction (cf.~Sections \ref{sec:eft} 
and \ref{sec:eft_future}), e.g., for processes that explore energies
approaching the breakdown scale of chiral EFT 
\cite{Lonardoni:2015sr,Madeira:2018yu,Cruz-Torres:2019tn,Lynn:2020xe}. 

A typical ansatz for a QMC trial state is
\begin{equation}
  \ket{\Phi_T} \equiv \mathcal{F}(\vec{a})\ket{\Phi(\vec{b})}\,,
\end{equation}
where $\mathcal{F}(\vec{a})$ is an operator that explicitly imprints correlations
on the mean-field like state $\ket{\Phi(\boldsymbol{b})}$, and 
$\boldsymbol{a,b}$ are vectors of tunable parameters. The first step of
most QMC calculations is a variational minimization of the energy in the
trial state\,,
\begin{equation}
  \min_{\vec{a}, \vec{b}} \frac{\matrixe{\Phi_T}{\HO}{\Phi_T}}{\braketn{\Phi_T}} \geq E_0\,,
\end{equation}
followed by an imaginary-time evolution to project out the true ground
state in a quasi-exact fashion:
\begin{equation}\label{eq:im_time}
  \ket{\Psi_0} \propto \lim_{\tau\to\infty} e^{-(\HO-E_T)\tau}\ket{\Phi_T}\,.
\end{equation}
This projection can be implemented using Monte Carlo techniques in a variety
of ways, which gives rise to different approaches like Green's Function 
Monte Carlo (GFMC) or Auxiliary-Field Diffusion Monte Carlo (AFDMC) 
\cite{Carlson:2015lq,Gandolfi:2020fv}. 

A major challenge in QMC calculations is that most commonly used algorithms 
suffer from some form of sign problem \cite{Carlson:2015lq,Gandolfi:2020fv}.
Many quantities of interest like the wave functions or local operator expectation 
values in these wave functions are not positive definite across their entire 
domain, which means that they cannot be immediately interpreted as probability 
distributions that the algorithms sample. This is one of the main reasons
why QMC methods can only be used with Hamiltonians that are either completely
local, or have a nonlocality that is at most quadratic in the momenta, e.g.,
$\vec{p}^2$ or $\vec{l}^2$.

While QMC applications in \emph{ab initio} nuclear structure have been 
focused on coordinate space, there \emph{are} a wide variety of approaches 
that merge QMC techniques with the configuration space approaches discussed 
in previous sections. Examples include sampling the intermediate-state 
summations in MBPT \cite{Hirata:2014vn}, diagrammatic expansions 
\cite{Prokofev:2007rp,Van-Houcke:2012wk,Scott:2019pt}, or the coefficients 
of correlated CC \cite{Roggero:2014we} or (No-Core) CI wave functions 
\cite{Otsuka:2001nx,Shimizu:2017ce,Booth:2009ct,Blunt:2019oe,Ten-no:2013ee}.

\subsubsection{Lattice Effective Field Theory}
\label{sec:left}

Lattice methods are nowadays widely used to simulate the dynamics of nonperturbative
field theories on finite space-time lattices. The most prominent example is Lattice
QCD, but implementations of various Effective Field Theories on the Lattice have 
been developed and applied with impressive outcomes in the past two decades ---
see, for example, Refs.~\cite{Nicholson:2017qi,Lee:2020qf,Lahde:2019tm,Lee:2017zj,Lee:2020qf}
and references therein, which also provide pedagogical introductions to Lattice
EFT for nuclear systems.

Lattice EFT simulations are built around the partition function, which is 
defined for a pure state $\ket{\Psi}$ as
\begin{equation}\label{eq:partition_tau}
  \mathcal Z(\tau) = \matrixe{\Psi(\tau=0)}{\exp\left(-H\tau\right)}{\Psi(\tau=0)}\,.
\end{equation}
Here, $\HO$ is 
an EFT Hamiltonian, typically truncated at a given order of the EFT's 
power counting scheme. In practice, the partition function is 
evaluated as a path integral in which field configurations are sampled
using Monte Carlo techniques.
At large $\tau$, one can extract information about the ground state 
and low-lying excited states of the system directly from $\mathcal{Z}$ 
(cf.~Section \ref{sec:qmc}), and general expectation values can be evaluated 
using
\begin{align}
  \expect{O}_\tau = \frac{1}{Z(\tau)}\dmatrixe{\Psi_0}{\exp(-H\tau /2) \,\OO\, \exp(-H\tau /2)}\,.
\end{align}

The use of discretized spatial lattices makes Lattice EFT particularly suited
for the description of nuclear states with complex geometries like cluster
structures \cite{Epelbaum:2012fv,Epelbaum:2014kx,Elhatisari:2017kq}. Depending on the size
of the lattices, it will also typically require less computational effort 
than the imaginary-time evolution of states that are formulated in
continuum coordinates, as in AFDMC or GFMC (see Section \ref{sec:qmc}).
Moreover, the development of the so-called adiabatic projection method (APM) 
\cite{Elhatisari:2015sf,Elhatisari:2019zi} in recent years has made it 
possible to compute scattering cross sections for reactions of (light) 
clusters on the lattice. Conceptually, the APM is reminiscent of the 
resonating-group method used to describe reactions in the NCSMC framework 
discussed in Section \ref{sec:ncsm}. 

Of course, Lattice EFT is not free of disadvantages, which are usually
caused by the discretization of space(time). The finite size and
lattice spacing are related to infrared (long-range, low-momentum) and 
ultraviolet (short-range, high-momentum) cutoffs of a calculation, which
need to be carefully considered. Since the recognition of cutoff scales
is an inherent aspect of EFTs, one can systematically correct for these
effects \cite{Klein:2018ul,Klein:2018vl}. The discrete lattice also breaks
continuous spatial symmetries that may need to be restored approximately
or exactly before comparisons with experimental data are made \cite{Nicholson:2017qi,Klein:2018vl}.

\section{The Past is Prologue: Achievements in the Last Decade}
\label{sec:current}

In this section, I will discuss selected achievements of the \emph{ab initio} 
nuclear many-body community in the past decade, and the issues that were 
encountered in the process. As stated in the introduction, this selection
is subjective, and giving full justice to the breadth of research accomplishments
is beyond the scope of this work. I hope that the present discussion will 
serve as an invitation for further exploration, for which the cited literature
may serve as a useful starting point.

\subsection{Benchmarking Nuclear Forces}
\label{sec:benchmark}
One of the biggest issues in nuclear theory was the lack of comparability
between different approaches for describing the structure of medium-mass or
heavy nuclei. These nuclei were well in reach of the Shell Model and nuclear
Density Functional Theory (DFT), but whenever issues emerged, it was unclear 
whether they resulted from approximations in the many-body method, or deficiencies 
in the effective interactions, i.e., the valence-space Hamiltonians or energy 
density functionals (EDF). Moreover, one cannot simply perform a valence CI
calculation with an EDF, or a DFT calculation with a Shell Model interaction,
because the interactions are tailored to their specific many-body method. 

The development of the RG/EFT and many-body methods discussed in Section \ref{sec:ingredients}
has opened up a new era for benchmarking the same nuclear interactions across 
multiple approaches, and on top of that, these methods provide a systematic 
framework for analyzing, and eventually quantifying, the reasons for differences
between the obtained results.

\begin{figure}[t]
  \setlength{\unitlength}{\textwidth}
  \begin{center}
    \includegraphics[width=0.7\unitlength]{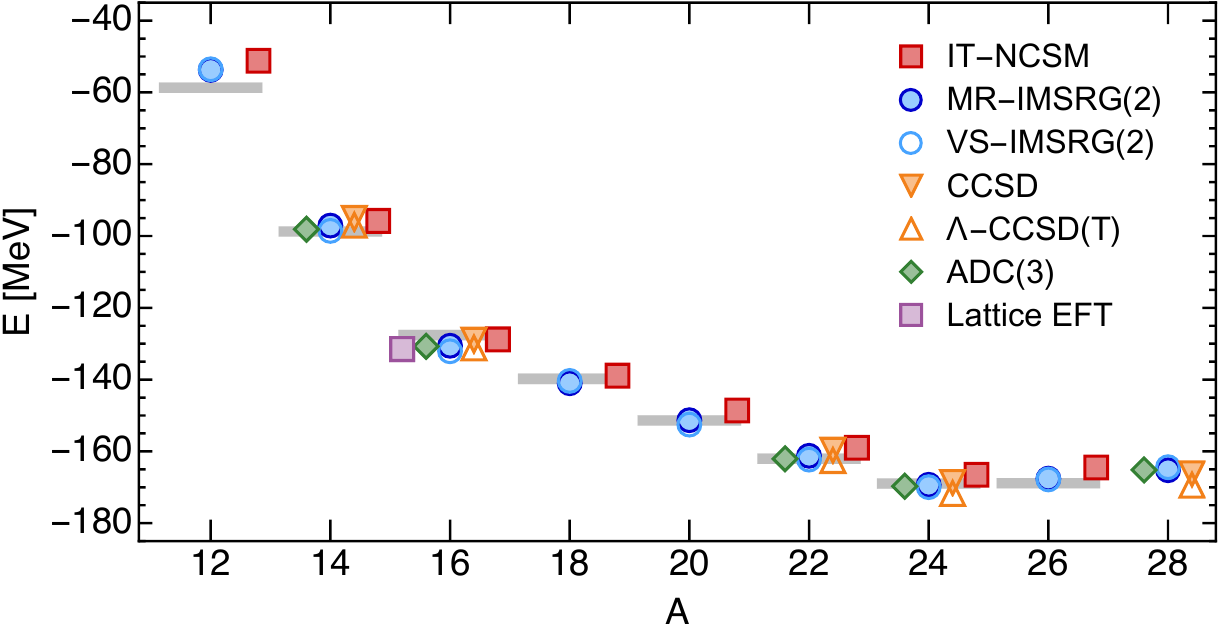}
  \end{center}
  \caption{Ground-state energies of the oxygen isotopes for various many-body approaches, 
    using the chiral NN+3N(400) interaction at $\lambdaSRG=1.88\fmi$ \cite{Roth:2011kx}.
    Details on the Lattice EFT calculation can 
    be found in Ref.~\cite{Epelbaum:2014kx}. Gray bars indicate experimental data 
    \cite{Huang:2017db}.
    }
  \label{fig:benchmark_O}
\end{figure}

One of the earliest testing grounds for \emph{ab initio} calculations of medium-mass
nuclei was the oxygen isotopic chain, which was accessible to all of the approaches 
that emerged at the beginning of the past decade. Figure \ref{fig:benchmark_O} shows 
the ground-state energies of even oxygen isotopes for the same chiral NN+3N interaction, 
obtained with several of the configuration space approaches introduced in 
Section \ref{sec:many_body}. In addition, results for applying various types of MBPT to
the same interaction and nuclei are presented in Ref.~\cite{Tichai:2020ft} --- I only 
refrained from including them here to avoid overloading the figure. As we can see,
the ground-state energies obtained from the different approaches are in good agreement
with each other and with experiment. Since our results include quasi-exact IT-NCSM
values, the deviation of the other methods' energies from these values provide us with
an estimate of the theoretical uncertainties due to any employed truncations, which is
on the order of 1-2\%.  As we can see from Fig.~\ref{fig:benchmark_O}, essentially all 
of the used many-body methods place the drip line in the oxygen isotopic chain at $\nuc{O}{24}$, 
although the signal is exaggerated. Continuum effects that have been omitted in these 
calculations would lower the energy of the $\nuc{O}{26}$ resonance, which is experimentally 
constrained to be a mere $18(7)\,\keV$ above the two-neutron threshold \cite{Kondo:2016fr},
and produce a very flat trend in the energies towards $\nuc{O}{28}$. Similar features
were found in calculations for other isotopic chains and other chiral interactions
\cite{Hergert:2014vn,Soma:2014fu,Holt:2019zf,Soma:2020dw}. The $\nuc{O}{16}$ ground
state energies obtained for the employed chiral NN+3N Hamiltonian are also compatible 
with a Lattice EFT result that was obtained at a similar resolution scale \cite{Epelbaum:2014kx}. 

This last comparison shows that some obstacles to the ideal cross-validation scenario
still remain. Since coordinate-space approaches like Lattice EFT or QMC are truly 
complementary to configuration-space methods, it would be highly desirable to test the 
same chiral NN+3N Hamiltonians in both types of calculations. However, the Hamiltonians 
used in configuration space are typically given in terms of harmonic oscillator matrix 
elements (especially if SRG evolved) instead of the coordinate-space operators required by 
Lattice EFT or QMC calculations. Furthermore, Lattice EFT and QMC
cannot handle all possible types of nonlocality in the Hamiltonian (cf.~Section \ref{sec:qmc}), 
including the forms generated by the nonlocal regulators that are favored for 
configuration-space Hamiltonians. Conversely, local chiral interactions 
that have been constructed explicitly for QMC applications \cite{Gezerlis:2014zr,Lonardoni:2018sz,Piarulli:2018xi,Lonardoni:2020yo,Piarulli:2020dp,Gandolfi:2020fv} exhibit slow model-space convergence 
in configuration-space calculations because they still tend to require a significant
repulsive core at short distance to describe nucleon-nucleon scattering data, albeit
a far weaker one than interactions like Argonne V18 \cite{Wiringa:1995or}.

\subsection{Extending the Reach of \emph{Ab Initio} Theory}

The reach of \emph{ab initio} 
many-body theory has increased dramatically over the past decade. Figure \ref{fig:nuclear chart}
illustrates this growing coverage of the nuclear chart, but it tells only part of the
story. The expansion has happened in many ``dimensions'' besides the mass number $A$, 
namely by pushing towards exotic nuclei via improved treatments of the continuum degrees of
freedom, filling in gaps in the coverage that are occupied by doubly open-shell nuclei
with strong intrinsic deformation, and expanding the types of observables that can
be computed from first principles. Recalling Section \ref{sec:benchmark}, the ongoing
push against the limitations of our many-body approaches will continue to grow the
opportunities for benchmarking current- and next-generation chiral Hamiltonians.

\subsubsection{Pushing the Mass Boundaries}

First calculations for selected nuclei and semi-magic isotopic chains up to tin were 
already published in the first half of the last decade \cite{Binder:2014fk,Hergert:2014vn,Soma:2014eu}.
For the most part, they were using a family of chiral NN+3N interactions that gave a
good description of the oxygen ground-state energies (cf.~Fig.~\ref{fig:benchmark_O}) 
as well as the spectroscopy of the lower $sd$-shell region \cite{Bogner:2014tg,Jansen:2014qf}.
However, the same interactions underpredict nuclear charge radii \cite{Lapoux:2016xu},
and start to overbind as we approached the calcium chain (cf.~Fig.~\ref{fig:ca_energies}), 
eventually leading to an overbinding of 1 MeV per nucleon in tin. While model-space convergence in CC, IMSRG 
and SCGF calculations suggested that calculations for heavier nuclei would have been 
technically possible, it made little sense to pursue them.

The growing number of results for medium-mass nuclei and the problems they revealed
motivated a new wave of efforts to refine chiral interactions. One direction of research
aimed to achieve a simultaneous description of nuclear energies and radii up to $\nuc{Ca}{48}$
by including selected many-body data in the optimization protocol of the chiral LECs.
This work resulted in the so-called \NNLOsat{} interaction \cite{Ekstrom:2015fk}. 
While \NNLOsat{} definitely improved radii \cite{Garcia-Ruiz:2016fk}, its model-space 
convergence was found to become problematically slow already in lower $pf$-shell 
nuclei \cite{Hagen:2016xe,Leistenschneider:2018mh,Soma:2020dw}. 

\begin{figure}[t]
  \setlength{\unitlength}{\textwidth}
  \begin{center}
    \includegraphics[width=0.5\unitlength]{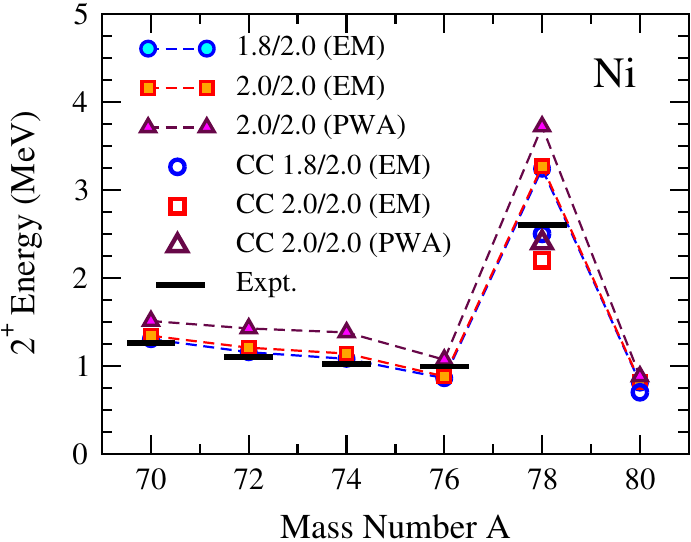}
  \end{center}
  \caption{
    Energies of the first excited $2^+$ states from VS-IMSRG \cite{Taniuchi:2019yq} and Equation-of-Motion CC \cite{Hagen:2016xe} calculations for several chiral two- plus three-nucleon interactions. Experimental values \cite{NNDC,Taniuchi:2019yq} are indicated as black bars. Data courtesy of J.~D.~Holt, J.~Men\'{e}ndez, and G.~Hagen.
  }
  \label{fig:Ni_isotopes}
\end{figure}

Simultaneously with
the efforts to develop new interactions, attention also turned towards an older, less 
consistently constructed family of chiral NN+3N interactions that exhibited reasonable
saturation properties in nuclear matter calculations \cite{Nogga:2004il,Hebeler:2011dq}. 
These forces are referred to as 
EM$\lambdaSRG/\Lambda$, where $\lambdaSRG$ indicates the resolution scale of
the NN interaction, the SRG-evolved \NNNLO{} potential of Entem and Machleidt 
\cite{Entem:2003th}, and $\Lambda$ is the cutoff of an \NNLO{} three-nucleon 
interaction whose low-energy constants have been adjusted to fit the triton 
binding energy and $\nuc{He}{4}$ charge radius \cite{Nogga:2004il,Hebeler:2011dq}.
In CC calculations for the nickel isotopes, Hagen et al. demonstrated that 
the EM1.8/2.0 interaction, in particular, allowed a good description of the 
energies of nuclei in the vicinity of $\nuc{Ni}{78}$ \cite{Hagen:2016xe}. 
As shown in Fig.~\ref{fig:Ni_isotopes}, these findings have been reinforced 
by subsequent VS-IMSRG calculations, as well as the experimental observation 
of the first excited $2^+$ state in this nucleus \cite{Taniuchi:2019yq}.

Since this initial application in medium-mass nuclei, the EM$\lambda/\Lambda$
family has seen widespread use in \emph{ab initio} calculations due to its
empirical quality, although the Hamiltonian's theoretical uncertainties  
are less well defined than for interactions that obey the chiral
power counting more rigorously. Indeed the EM1.8/2.0 interaction was used in VS-IMSRG
calculations to produce what is to my knowledge the first attempt at producing an 
\emph{ab initio} mass table for nuclei up to the iron isotopes \cite{Holt:2019zf}.
For selected nuclei up to the tin region, it also yields converged energies 
for ground and low-lying states that are in good agreement with experimental
data \cite{Simonis:2017hl,Morris:2018hw}. It also yields slightly larger
radii than previous interactions, although the underprediction is
not eliminated entirely (see Refs. \cite{Garcia-Ruiz:2016fk,Simonis:2017hl} 
and Section \ref{sec:obs}). 

Multiple applications of the EM$\lambda/\Lambda$ Hamiltonians in support of 
spectroscopy experiments have been published in recent years
(see, e.g., \cite{Henderson:2018la,Leistenschneider:2018mh,Evitts:2019kh,Liu:2019if,Xu:2019bx}), 
and additional studies are underway, including an effort to better understand 
what makes the EM1.8/2.0 Hamiltonian so successful. Furthermore, a new generation 
of chiral NN+3N interactions is now available for applications in medium-mass and 
heavy nuclei \cite{Reinert:2018uq,Huther:2019yz,Drischler:2019qf,Hoppe:2019th,Soma:2020dw}.

\subsubsection{Towards the Drip Lines}
\label{sec:drip}

Neutron-rich nuclei are excellent laboratories for disentangling the interplay
of nuclear interactions, many-body correlations and the continuum. Thus, data 
from the experimental push towards the drip line can offer important constraints
for the refinement of chiral interactions if the many-body truncations and
continuum effects are under control.

In practice, \emph{ab initio} results for observables like the absolute energies 
of states still exhibit significant scale and scheme dependence due to truncations
that are made in the EFT, the potential implementation of SRG evolutions, and the
many-body methods. Since such variations tend to be systematic within families of
interactions (and sometimes even across multiple families), differential quantities
like separation and excitation energies or transition matrix elements often exhibit 
a weaker scale and scheme dependence --- note, for example, the small systematic
variation of the first excited $2^+$ states of the neutron-rich nickel isotopes for 
EM$\lambda/\Lambda$ interactions. This makes energy differences an ideal observable for 
confronting \emph{ab initio} results with experimental data.

\begin{figure}[t]
  \setlength{\unitlength}{\textwidth}
  \begin{center}
    \includegraphics[width=\unitlength]{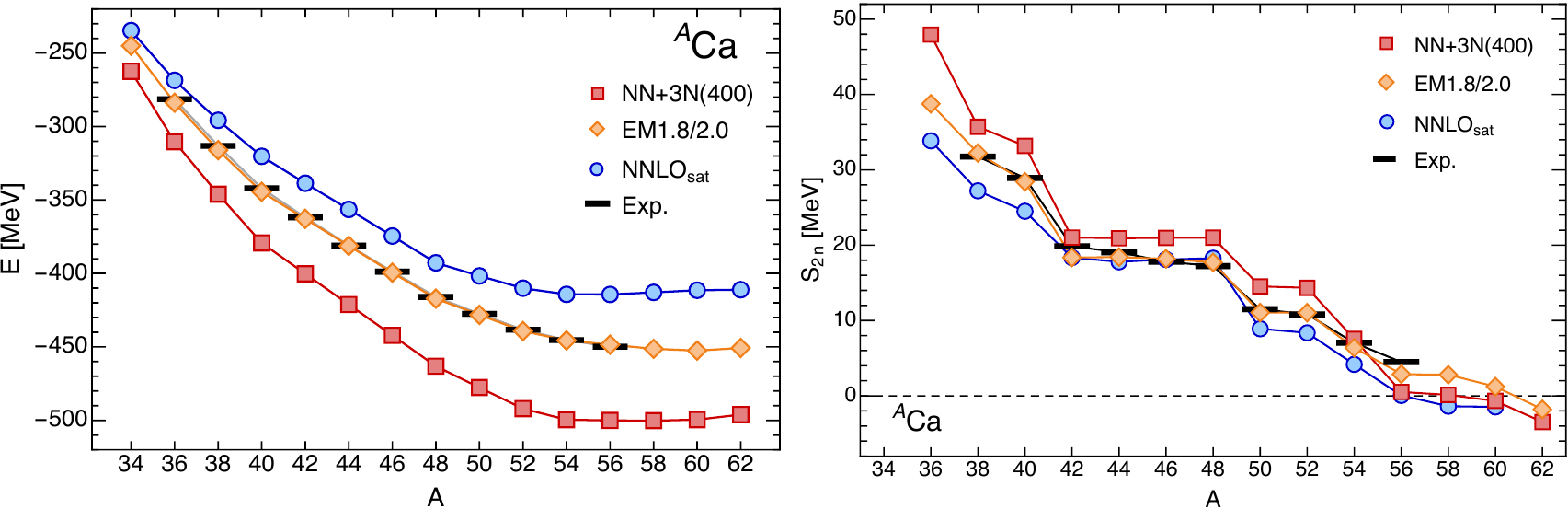}
  \end{center}
  \caption{
    Ground-state and two-neutron separation energies for several chiral NN+3N interactions
    from MR-IMSRG(2) calculations. Experimental data are indicated by black bars \cite{Huang:2017db,Michimasa:2018qe}.
  }
  \label{fig:ca_energies}
\end{figure}

Let us consider two-neutron separation energies as a concrete example. Sudden 
drops in these observables are a signal of (sub)shell closures (albeit not 
universally \cite{Garcia-Ruiz:2016fk}) and in the neutron-rich domain, they are 
important indicators for the proximity of the drip line. Figure~\ref{fig:ca_energies} 
shows MR-IMSRG ground-state and two-neutron separation energies of the calcium
isotopes, obtained with the NN+3N(400) interaction used in 
Fig.~\ref{fig:benchmark_O}, as well as the \NNLOsat{} and EM1.8/2.0 interactions 
briefly discussed in the previous section. We note the overbinding produced by 
NN+3N(400) and the baffling accuracy of the EM1.8/2.0 results, given the 
approximations that went into the construction of this force, as well as the 
MR-IMSRG truncation. Common to all three interactions is the emergence of a 
very flat trend in the ground-state and separation energies in neutron-rich 
calcium isotopes, which will likely be further enhanced by the inclusion of 
continuum effects, and extended beyond the shown mass range. Similar flat trends 
emerge in many isotopic chains, as shown both in \emph{ab initio} surveys based
on chiral interactions \cite{Soma:2020dw,Soma:2020fj,Holt:2019zf} as well as 
a sophisticated Bayesian analysis of empirical EDF models \cite{Neufcourt:2019tx}. Naturally, 
this will make the precise determination of the neutron drip line in the medium-mass region 
a challenging task, but also suggests that interesting features like alternating 
patterns of unbound odd nuclei and weakly-bound even nuclei with multi-neutron 
halos could emerge. This is an exciting prospect for the experimental programs 
at rare-isotope facilities.

With the exception of the NCSMC and HORSE methods discussed in Section \ref{sec:ncsm}, the inclusion 
of continuum degrees in configuration-space techniques has been focused on
the use of the Berggren basis \cite{Berggren:1968hb}. While such calculations
are challenging due to the significantly increased single-particle basis size 
and the difficulties of handling the resulting complex symmetric Hamiltonians,
applications in CC (see Refs.~\cite{Hagen:2014ve,Hagen:2016rb} and references 
therein), both valence and No-Core Gamow Shell Model \cite{Papadimitriou:2013dt,Rotureau:2013km,Shin:2017ef,Sun:2017zm,Hu:2020rw}
and IMSRG \cite{Hu:2019kx} calculations have been published. Common to all these approaches
is that a configuration space interaction that is given in terms of SHO matrix
elements is expanded on a basis containing SHO and Berggren states, hence it
is still an open question how a direct implementation of the interactions in 
a basis with continuum degrees of freedom might modify existing results. It
is worth noting that such a construction has been achieved for phenomenological
GSM interactions that have been tuned for light nuclei \cite{Jaganathen:2017fq,Fossez:2017ty,Fossez:2017lw,Fossez:2018ec,
Wang:2019wj,Mao:2020wo}.

\begin{figure}[t]
  \setlength{\unitlength}{\textwidth}
  \begin{center}
    \includegraphics[width=0.9\unitlength]{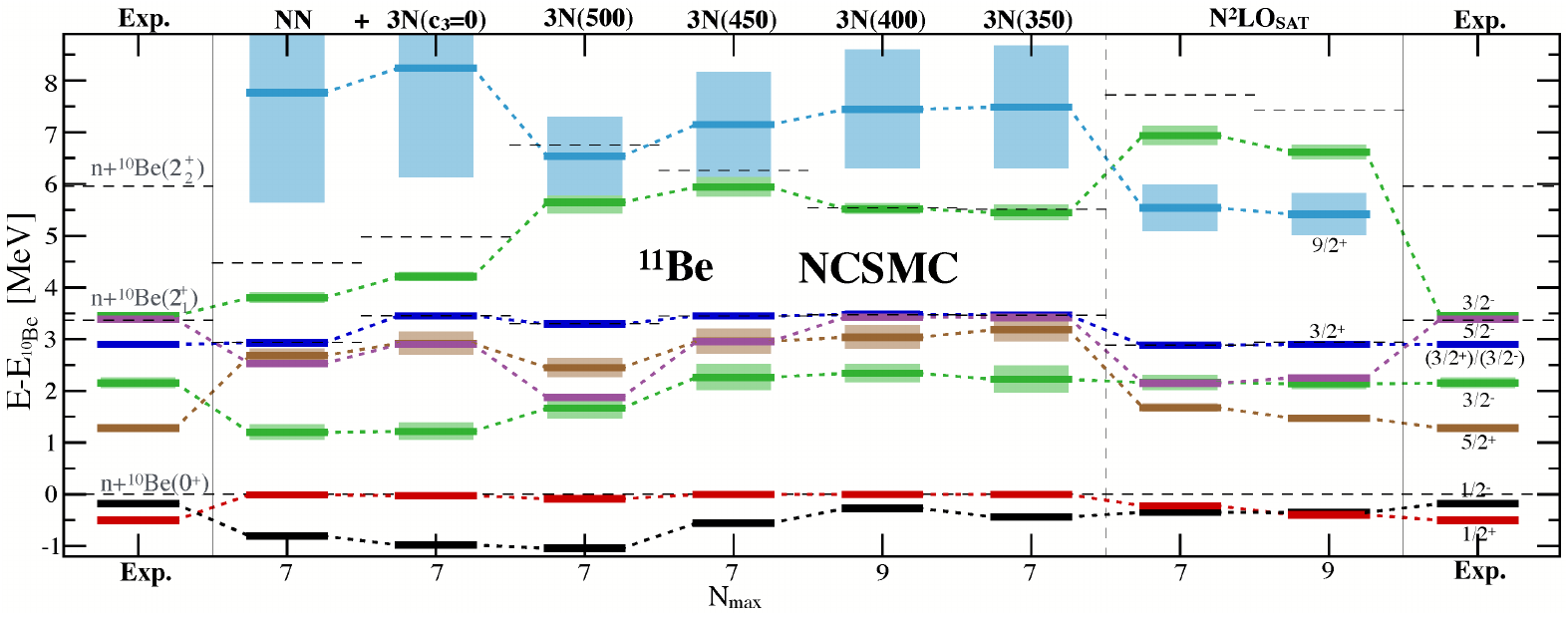}
  \end{center}
  \caption{
    NCSMC spectrum of $\nuc{Be}{11}$ with respect to the $n+\nuc{Be}{10}$ threshold. Dashed 
    black lines indicate the energies of the $\nuc{Be}{10}$ states. Light boxes indicate 
    resonance widths. See Ref.~\cite{Calci:2016kq} for details. Figure reprinted with 
    permission from the American Physical Society.
  }
  \label{fig:be11}
\end{figure}

In light nuclei, the NCSMC has been applied with impressive success to describe 
a variety of exotic nuclei with up to three-cluster structures. For example,
Calci et al.~\cite{Calci:2016kq} carried out NCSMC calculations for $\nuc{Be}{11}$ 
with several chiral NN+3N interactions to investigate the parity inversion of 
the ground and first-excited states in this nucleus from first principles. The 
authors found that the coupling between the NCSM and RGM sectors of the generalized
eigenvalue has strong effects, but that among the tested interactions, 
only \NNLOsat{} can produce the experimentally observed ordering of the states 
(see Fig.~\ref{fig:be11}). However, it still underpredicts the splitting of
these levels and as a result, overestimates the cross section for the 
photodisintegration $\nuc{Be}{11}(\gamma,n)\nuc{Be}{10}$. Additional applications
of the NCSMC for exotic nuclei can be found in the review \cite{Navratil:2016rw}
and references therein, as well as the more recent works \cite{Kumar:2017dp,Vorabbi:2018rp,Vorabbi:2019rr}.

\subsubsection{Accessing More Observables}
\label{sec:obs}

The capabilities of \emph{ab initio} approaches have also significantly expanded
when it comes to the evaluation of observables other than the energies. 

\begin{figure}[t]
  \setlength{\unitlength}{0.48\textwidth}
  \begin{center}
    \includegraphics[width=2\unitlength]{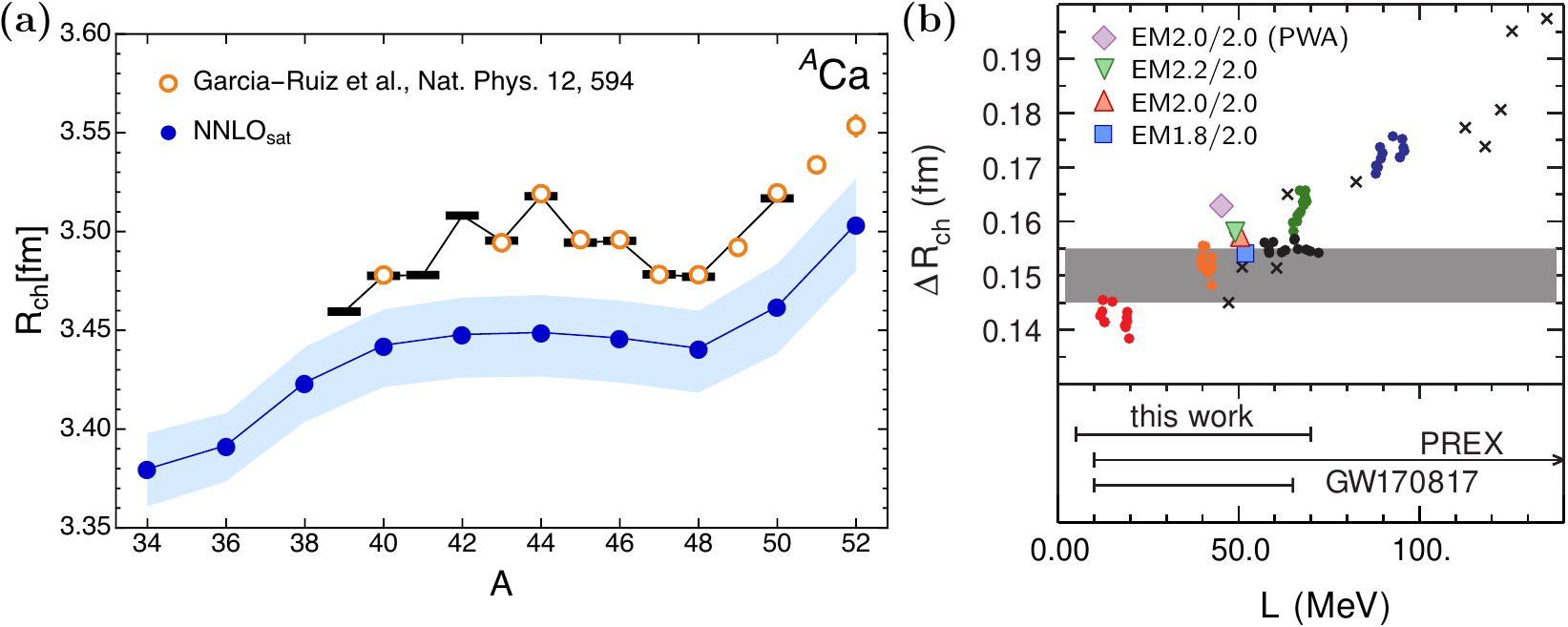}
  \end{center}
  \caption{Panel (\textbf{a}) Calcium charge radii from MR-IMSRG(2) calculations with \NNLOsat{} 
    . The shaded area indicates uncertainties from basis convergence. Black bars and orange circles 
    indicate experimental data \cite{Angeli:2013rz,Garcia-Ruiz:2016fk}.
    Panel (\textbf{b}): Mirror charge radius difference of $\nuc{Ca}{36}$ and $\nuc{S}{36}$ 
    versus the slope of the symmetry energy, $L$, at nuclear saturation, for the EM$\lambda/\Lambda$
    interactions (symbols as indicated in the legend), compared to Skyrme functionals (solid circles)
    and Relativistic Mean Field models (crosses). The band indicates the experimental result from the 
    BECOLA facility at NSCL. See Ref.~\cite{Brown:2020dz} for details.
  }
  \label{fig:rch_Ca}
\end{figure}

\noindent \textbf{Nuclear Radii.} Figure \ref{fig:rch_Ca}
shows MR-IMSRG results for the charge radii of calcium isotopes. The left panel
illustrates the reasonable reproduction of the $\nuc{Ca}{40}$ and $\nuc{Ca}{48}$
charge radii that can be obtained for \NNLOsat{}. The MR-IMSRG(2) results
are slightly smaller than the experimental data due to differences in the
truncations from the CCSD charge radius calculations that were used in the \NNLOsat{}
optimization protocol \cite{Ekstrom:2015fk}. 
Note the steep
increase in the experimental charge radii beyond $\nuc{Ca}{48}$: At the time of
the measurement, \NNLOsat{} was the only chiral NN+3N interaction exhibiting 
this feature, although other more recent interactions can replicate this trend
as well \cite{Soma:2020dw,Soma:2020fj}. Also note that none of the calculations
are able to reproduce the inverted arc of the charge radii between $\nuc{Ca}{40}$
and $\nuc{Ca}{48}$. In a CI picture, it is caused by strong mixing with 4p4h 
excitations into the $pf$-shell \cite{Caurier:2001oq}. Since the MR-IMSRG(2) 
calculations shown here included only up to (generalized) 2p2h excitations and 
used particle-number projected Hartree-Fock Bogoliubov vacua as reference states 
that do not contain collective correlations (cf.~Section \ref{sec:imsrg}), it is not 
surprising that the inverted arc cannot be reproduced. We will return to this issue 
of missing collectivity later.  

While the EM$\lambda/\Lambda$ interactions underpredict the absolute charge 
radii, they fare quite well in the description of radius differences, as 
suggested in the previous section. Figure \ref{fig:rch_Ca}(b) is adapted from
a recent study that suggests a correlation between the charge radius difference
of mirror nuclei, $\Delta R_\text{ch}$, and the slope of the symmetry energy in 
the nuclear matter equation of state \cite{Brown:2020dz}. We see that the 
MR-IMSRG results for $\Delta R_\text{ch}$ are actually compatible with results
from a multitude of Skyrme EDFs, and the value for the magic EM1.8/2.0 interaction 
falls into the uncertainty band of the experimental result.

\noindent \textbf{Electromagnetic Transitions.}
Since the second half of the past decade, \emph{ab initio} calculations for 
transitions in medium-mass nuclei have become more frequent, owing to the 
appropriate extensions of the IMSRG, CC and SCGF methods 
\cite{Parzuchowski:2017ta,Henderson:2018la,Raimondi:2019wj}.
While results for transitions that are dominated by a few nucleons, e.g., 
$M1$ transitions \cite{Parzuchowski:2017ta} or $\beta$ decays (see Ref.~\cite{Gysbers:2019df}
and the discussion below) can be quite good, the description of collective 
transitions is hampered by inherent truncations of these many-body methods,
which are better suited for dynamical, few-particle correlations (see Sections \ref{sec:imsrg}
and \ref{sec:cc}). Results from the SA-NCSM \cite{Launey:2016ef,Dytrych:2020db}
and the IM-GCM discussed in 
Section \ref{sec:imsrg} show that the modern chiral interactions themselves
adequately support the emergence of nuclear collectivity. 

\begin{figure}[t]
  \setlength{\unitlength}{\textwidth}
  \begin{center}
    \includegraphics[width=0.85\unitlength]{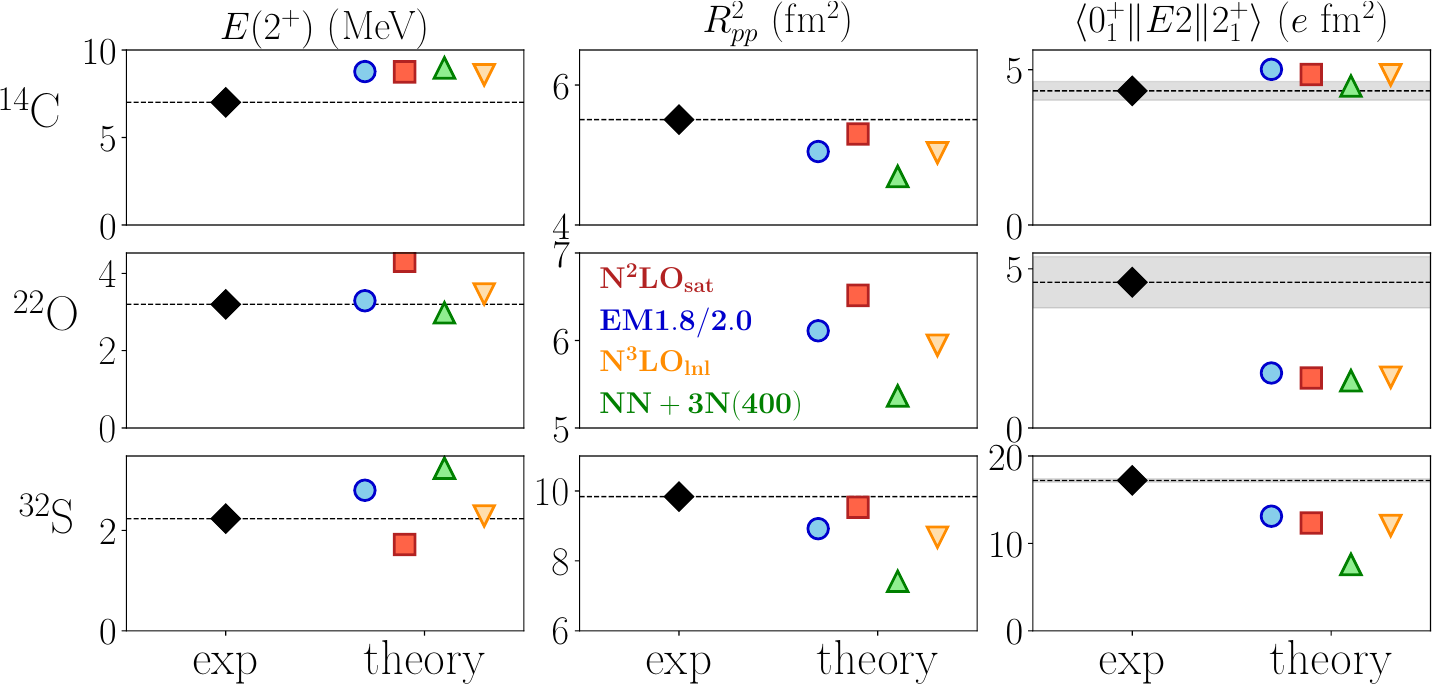}
  \end{center}
  \caption{
    Energies of the first excited $2^+$ state, proton mean square radius and quadrupole
    transition matrix elements for selected nuclei, based on VS-IMSRG(2) calculations with
    multiple chiral NN+3N interactions. See Refs.~\cite{Parzuchowski:2017ta} and
    \cite{Stroberg:2019th} for more details. Experimental values (with uncertainties indicated
    by bands) are taken from
    \cite{Angeli:2013rz,Pritychenko:2016ys}. Figure courtesy of R. Stroberg.
  }
  \label{fig:transitions}
\end{figure}

Consider for example Fig.~\ref{fig:transitions}, which shows VS-IMSRG(2) results for
the quadrupole transition from the first excited $2^+$ state to the ground 
state in $\nuc{C}{14}$, $\nuc{O}{22}$ and $\nuc{S}{32}$ \cite{Parzuchowski:2017ta}. The picture is fairly
consistent for all four chiral NN+3N interactions that were used in the study:
The $2^+$ energies are described quite well, but energies are not very sensitive
to the details of the nuclear wave functions. In $\nuc{C}{14}$, the $E2$
transition is weakly collective, so the $E2$ matrix element is reasonably reproduced,
while the matrix element for the collective transition in $\nuc{S}{32}$ is underpredicted
by 25-50\%. The NN+3N(400) interaction gives a particularly poor result, but this 
is also related to the significant underestimation of the point-proton radius we
obtain for this Hamiltonian, as discussed earlier. 

The result for $\nuc{O}{22}$ deserves special attention. The $E2$ transition matrix 
element is only a third of the experimental value, although the transition 
is only weakly collective. However, $\nuc{O}{22}$ only has neutrons in an $sd$
valence space, so the $E2$ matrix element would vanish in a conventional Shell Model 
calculation unless the neutrons have an effective charge. Such effective charges
must be introduced by hand and fit to data in phenomenological Shell Model calculations.
Here, we see that the VS-IMSRG decoupling naturally induces a non-vanishing quadrupole
moment through an effective neutron charge in the one-body transition operator as well
as an induced two-body contribution (see Ref.~\cite{Parzuchowski:2017ta}, and 
Ref.~\cite{Raimondi:2019wj} for an analogous effort in SCGF theory). It is likely that 
the $E2$ strength could be improved by performing the VS-IMSRG calculation in
a $psd$ valence space, so that the proton dynamics is treated explicitly instead of 
implicitly by valence-space decoupling. Until recently, we were unable to perform such
a multi-shell decoupling because of the IMSRG version of the intruder-state problem, 
but a promising workaround was introduced in Ref.~\cite{Miyagi:2020rm}.

\noindent \textbf{Gamow-Teller Transitions.}
In recent years, there have been concerted efforts to understand the mechanisms 
behind the empirically observed quenching of Gamow-Teller (GT) transitions in 
medium-mass nuclei, in part due to its relevance to neutrinoless double-beta
decay searches (see below). In Ref.~\cite{Gysbers:2019df}, the authors show that 
this issue 
is largely resolved by properly accounting for the scale and scheme dependence of 
configuration-space calculations. By dialing the resolution scale to typical values 
favored by approaches like NCSM, CC and VS-IMSRG, correlations are shifted from the 
wave functions into induced two- and higher-body contributions to the renormalized 
transition operator, just as in the quadrupole case discussed above.

The transition operator, including two-body currents, is consistently evolved
to lower resolution scale alongside the nuclear interactions, keeping the induced
contributions. The transition matrix elements of the evolved operator are 
then computed with the NCSM in light nuclei, and VS-IMSRG in $sd$- and $pf$-shell
nuclei, leading to agreement with experimental GT strengths within a few \%.
In contrast, the bare GT operator must be quenched by 20-25\% via the introduction
of an effective axial coupling, $g^\text{eff}_{A} < g_A$, to yield 
agreement with experimental beta decay rates.

The GT transitions in light nuclei have also been evaluated in the GFMC, most
recently with consistently constructed local chiral interactions and currents
\cite{Pastore:2018et,King:2020cd}. Interestingly, the inclusion of two-body
currents seems to consistently enhance the GT matrix elements, while it
tends to quench the matrix element in NCSM calculations. Since this is almost
certainly related to the differences in the resolution scale and calculation
scheme, the disentanglement of these observables might yield further
insights into the interplay of wave function correlations and the renormalization
of the transition operators.

\begin{figure}[t]
  \setlength{\unitlength}{\textwidth}
  \begin{center}
    \includegraphics[width=\unitlength]{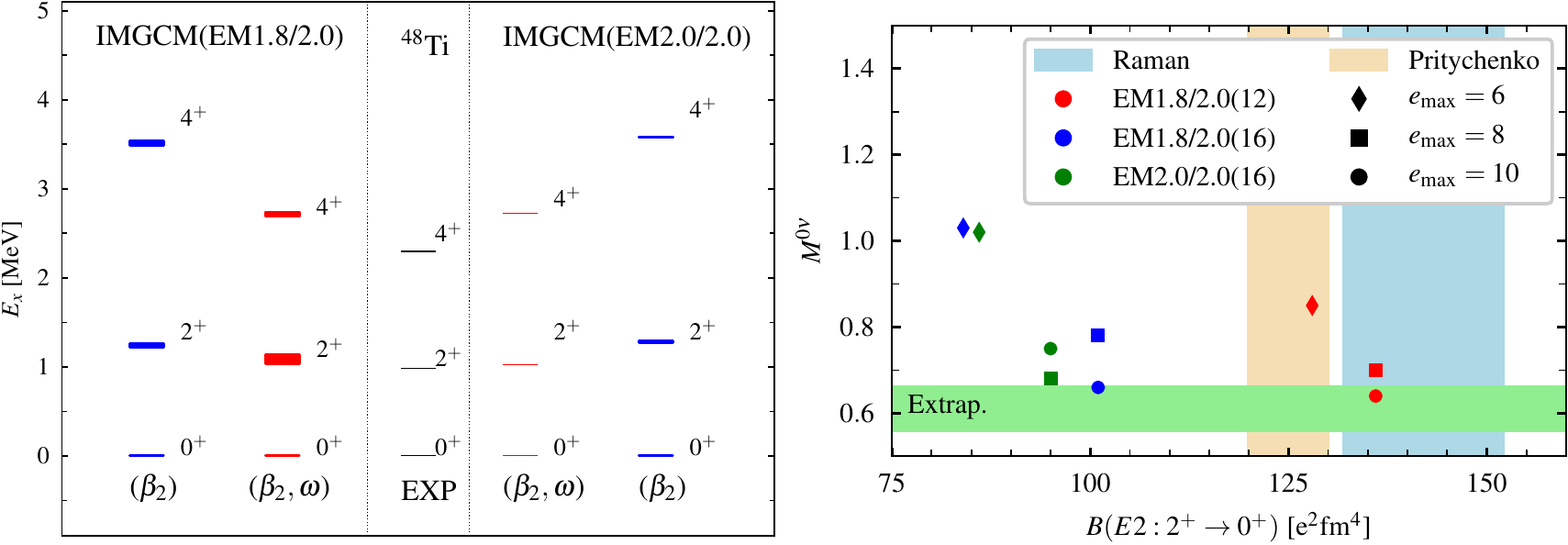}
  \end{center}
  \caption{
    IM-GCM description of the neutrinoless double beta decay $\nuc{Ca}{48}\to\nuc{Ti}{48}$,
    using EM$\lambda/\Lambda$ interactions: Low-lying spectrum of $\nuc{Ti}{48}$ and its
    compression through the admixing of cranked configurations (\textbf{a}) and the
    nuclear matrix element vs. B(E2) transition probability (\textbf{b}). See text 
    and Ref.~\cite{Yao:2020mw} for details.
    Panel (\textbf{a}) courtesy of J.~M.~Yao, panel \textbf{(b)} reprinted with permission 
    from the American Physical Society.
  }
  \label{fig:imgcm}
\end{figure}

\noindent \textbf{Neutrinoless Double Beta Decay.} Due to the high impact the 
observation of neutrinoless double beta decay (or lack thereof) would have on
particle physics and cosmology, the computation of nuclear matrix elements (NMEs) 
for neutrinoless double beta decay is a high priority for nuclear structure 
theory. Precise knowledge of the NMEs for various candidate nuclei is required
to extract key observables like the absolute neutrino mass scale from the measured
lifetimes (or at least, any new bounds that would be provided by experiment).
Most calculations of the NME to date were subject to the lack of comparability 
between phenomenological nuclear structure results that was discussed in Section \ref{sec:benchmark},
hence a new generation of \emph{ab initio} calculations with quantified uncertainties
is required.

A major step in that direction was the first calculation of the NME for the decay
$\nuc{Ca}{48}\to\nuc{Ti}{48}$ based on chiral interactions \cite{Yao:2020mw}. The
IM-GCM approach discussed in Section \ref{sec:imsrg} was used to describe the structure
of the intrinsically deformed daughter nucleus $\nuc{Ti}{48}$, achieving a satisfactory
reproduction of the low-lying states and their quadrupole transitions (see Fig.~\ref{fig:imgcm}).
Since the initial publication (blue spectra in Fig.~\ref{fig:imgcm}(a)), the 
description of the excited states has been improved further through the admixing
of cranked configurations (red spectra), without affecting the NME (Fig.~\ref{fig:imgcm}(b)).
Work on quantifying the uncertainties due to the many-body method, the Hamiltonian,
and the transition operator is underway, in preparation for the computation of the
NMEs of more realistic candidate nuclei like $\nuc{Ge}{76}$ and $\nuc{Xe}{136}$.

\subsubsection{Response and Scattering}
\label{sec:response_and_scattering}

From the computation of transitions between low-lying levels, it is only a 
small step to the computation of nuclear response functions and cross sections, 
although the implementation can be challenging and the applications are often
computationally expensive.

\noindent\textbf{Nuclear Response Functions.}
In light nuclei, GFMC is a powerful yet numerically heavy 
tool for computing exact nuclear response functions (see, e.g., 
Refs.~\cite{Lovato:2014mi,Lovato:2018xr}). In medium-mass nuclei, applications
of SCGF and CC techniques to the computation of the nuclear response have
been published in recent years. As mentioned in Section \ref{sec:scfg}, the Green's 
functions computed in the standard or Gor'kov ADC Green's function schemes 
inherently contain information about the nuclear response that has been used 
to study both electromagnetic and weak processes of medium-mass nuclei 
\cite{Rocco:2018va,Barbieri:2019jw,Rocco:2019eq,Raimondi:2019nt,Rocco:2020zv}.

In the Coupled Cluster framework, response functions have been computed by 
merging CC with up to Triples excitations with the Lorentz Integral Transformation
(LIT) technique \cite{Bacca:2013lh,Bacca:2014ye,Miorelli:2016mm,Miorelli:2018pb,Simonis:2019qr}. 
Immediately after its inception, this approach was used to for the first \emph{ab initio} 
calculations of dipole response and the related photodisassociation cross 
section of medium-mass closed-shell nuclei \cite{Bacca:2013lh,Bacca:2014ye}.
More recently, it was used to compute the electric dipole polarizability $\alpha_D$
of nuclei like $\nuc{Ca}{48}$ \cite{Miorelli:2016mm,Birkhan:2017ye,Miorelli:2018pb}
and $\nuc{Ni}{68}$ \cite{Kaufmann:2020bh}. Together with measurements of the
charge radius, this quantity can be used to constrain \emph{ab initio} calculations
that will in turn allow the theoretical extraction of the neutron point
radius as well as the thickness of the neutron skin.

An important application for nuclear response calculations is to map out 
the neutrino response of $\nuc{Ar}{40}$, the primary target material in detectors
for the short-baseline \cite{Antonello:2015lea} and long-baseline neutrino experiments, like 
the Deep Underground Neutrino Experiment (DUNE) Far Detector \cite{Abi:2020kk,Abi:2020qa}. 
At low energies, the cross section for coherent neutrino elastic scattering is 
essentially determined by the weak form factor of $\nuc{Ar}{40}$, which has recently 
been computed using CC techniques \cite{Payne:2019fb}. This work is complementary
to SCGF calculations of the neutrino response in the region of the quasi-elastic 
peak by Barbieri et al.~\cite{Barbieri:2019jw}.

\begin{figure}[t]
  \setlength{\unitlength}{\textwidth}
  \begin{center}
    \includegraphics[width=\unitlength]{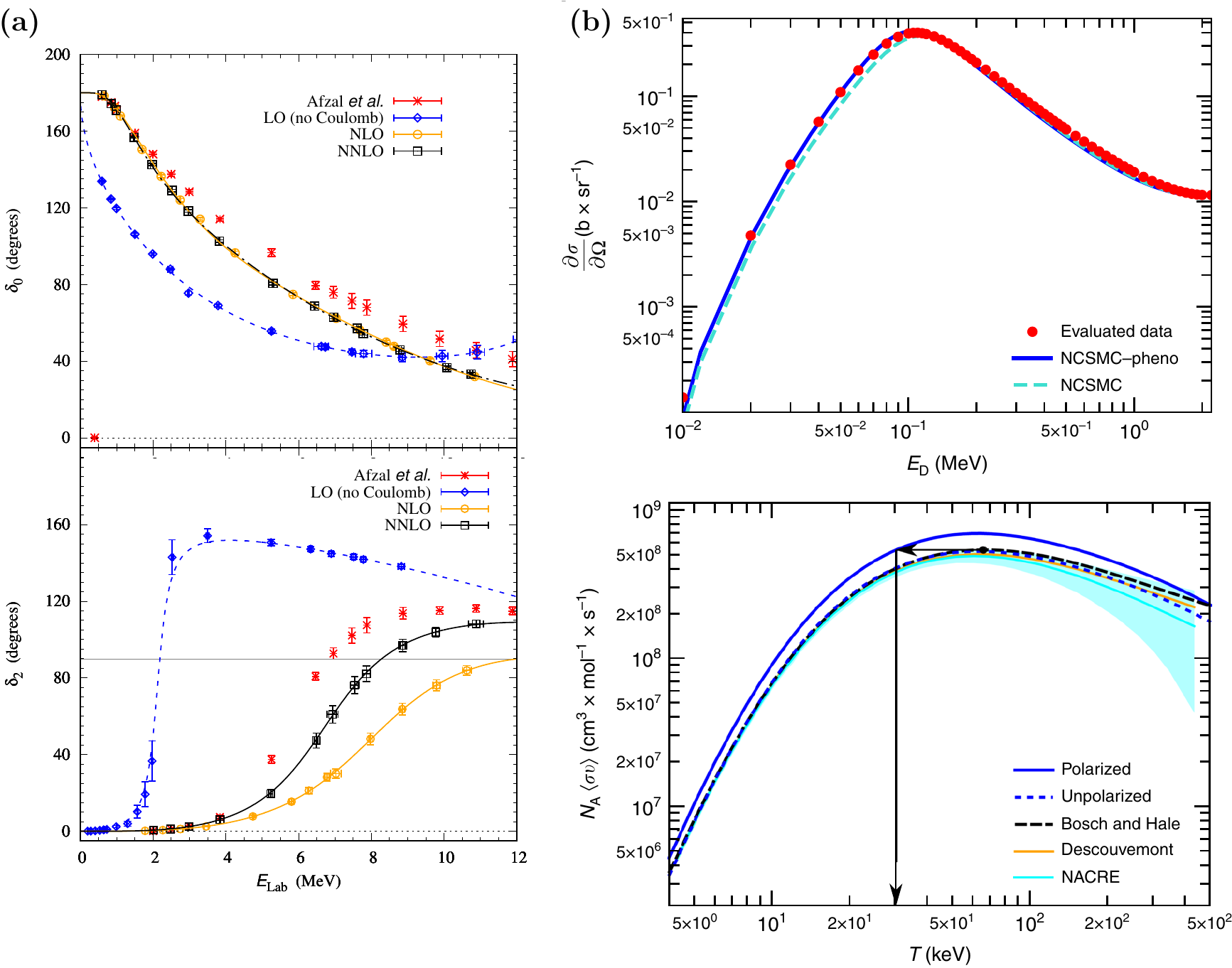}
  \end{center}  
  \caption{\emph{Ab initio} calculations of nuclear reactions.
    Panel \textbf{(a)}: $S$ ($\delta_0$) and $D$-wave phase shifts ($\delta_2$) for 
    $\alpha$-$\alpha$ scattering at various orders in Lattice EFT.
    For details, see \cite{Elhatisari:2015sf,Elhatisari:2019zi}. Figure courtesy of S.~Elhatisari.
    Panel \textbf{(b)}: NCSMC results for the deuterium-tritium (D--T) fusion cross section (top) 
    and reaction rate (bottom). The figure compares the rates for unpolarized and polarized fuel, 
    as well as rates obtained from widely adopted parametrization of the fusion cross section 
    (see Ref.~\cite{Hupin:2019oz} for details). The arrows are included to the guide the 
    reader's eye (see text). Figure reprinted from \cite{Hupin:2019oz} under a CC BY 4.0 license.
  }
  \label{fig:reactions}
\end{figure}

\noindent\textbf{Nuclear Reactions.}
As discussed in Section \ref{sec:many_body}, there has been enormous progress in the
development of unified treatments of \emph{ab initio} nuclear structure and reactions.
Here, I want to highlight two among a bevy of impressive results. Figure \ref{fig:reactions}(a)
shows $S-$ and $D-$wave phase shifts for $\alpha-\alpha$ scattering, computed 
order by order in Lattice EFT \cite{Elhatisari:2015sf,Elhatisari:2019zi}. These 
calculations are made possible by the lattice's capability to describe clustered
states (also see Refs.~\cite{Epelbaum:2012fv,Epelbaum:2014kx,Elhatisari:2017kq}), 
as well the development of the APM and associated algorithms. 
The results for the phase shifts show the desired order-by-order improvement, and 
the inclusion of higher-order terms of the chiral 
expansion is expected to improve agreement with experimental data. The near
identical \NLO{} and \NNLO{} phase shifts in the $S-$wave appear to be the
result of an accidental cancellation that is not occurring in the $D-$wave phase 
shifts.

In Ref.~\cite{Hupin:2019oz}, the authors studied deuterium-tritium (D-T) fusion using the 
NCSMC. One of the main results of this work is shown in Fig.~\ref{fig:reactions}(b), which 
compares the NCSMC D--T reaction rates for polarized and unpolarized fuels to each other, as 
well as rates obtained with several widely used parameterizations of the D--T fusion cross 
section. The NCSMC calculations indicate that for an experimentally realizable polarized fuel 
with aligned spins, a reaction rate of the same magnitude as for unpolarized fuel can be 
achieved at about half the temperature. Naturally, this suggests that polarized D-T fuels will 
allow a more efficient power generation in thermonuclear reactors.

\subsection{Emergence of Empirical Nuclear Structure Models from \emph{Ab Initio} Calculations}
\label{sec:lore}

The progress in \emph{ab initio} calculations over the past decade has not only 
led to impressive results for nuclear observables, but also revealed the long-surmised
underpinnings of empirical models of nuclear structure. In many cases, the ideas
that led to the formulation of such models were shown to be correct, but they
could not be verified at the time because RG and EFT techniques or sufficient
computing power for a more thorough exploration were not available.

\noindent
\textbf{The Nuclear Shell Model.} The first prominent example I want to discuss
is the nuclear Shell Model and some of the ``folklore'' 
surrounding it. We can immediately make the observation that the Shell model 
picture is inherently a low-momentum description of nuclear structure. It is
based on the assumption that nucleons are able to move (almost) independently
in a mean field potential, and that nuclear spectra can be explained by the 
mixing of a few valence configurations above an inert core via the residual 
interaction. As we know now, the existence of a bound mean-field solution and 
a weak, possibly perturbative residual interaction relies on the decoupling of
low and high momenta in the nuclear Hamiltonian \cite{Bogner:2010pq,Tichai:2016vl,Hoppe:2017fm},
e.g., by an SRG transformation. Historical approaches to exploit this connection 
to construct the Shell model from realistic nuclear forces \cite{Bertsch:1965zr,Kuo:1966ij,Kuo:1967qf} 
failed in part because the decoupling of the momentum scales via Brueckner's 
$G-$matrix formalism \cite{Brueckner:1954qf,Brueckner:1955rw,Day:1967zl} was not as 
good as believed \cite{Bogner:2010pq}.

In addition to the momentum-space decoupling, one must also decouple the valence
space configurations from the excluded space. This can be achieved using a variety
of techniques (cf.~Sections \ref{sec:imsrg}--\ref{sec:ncsm}), and either by performing 
transformations in sequence, or designing 
a single procedure that achieves both types of decoupling simultaneously. In practice, 
the former strategy tends to be more efficient and less prone to truncation errors  
--- an example is the VS-IMSRG decoupling of Hamiltonians that have been 
evolved to a low resolution scale by means of a prior SRG evolution (see Sections \ref{sec:srg}
and \ref{sec:imsrg}, as well as Ref.~\cite{Stroberg:2019th}). An added benefit
of using low-momentum interactions is that the Shell Model wave functions will
qualitatively resemble those obtained by a no-core method using the same Hamiltonian
without valence decoupling. This facilitates qualitative comparisons and allow us 
to apply the same intuitive picture. For quantitative comparisons, the effects of 
all unitary transformations must be carefully taken into account \cite{Duguet:2015lq}.

\begin{figure}[t]
  \setlength{\unitlength}{\textwidth}
  \begin{center}
    \includegraphics[width=0.8\unitlength]{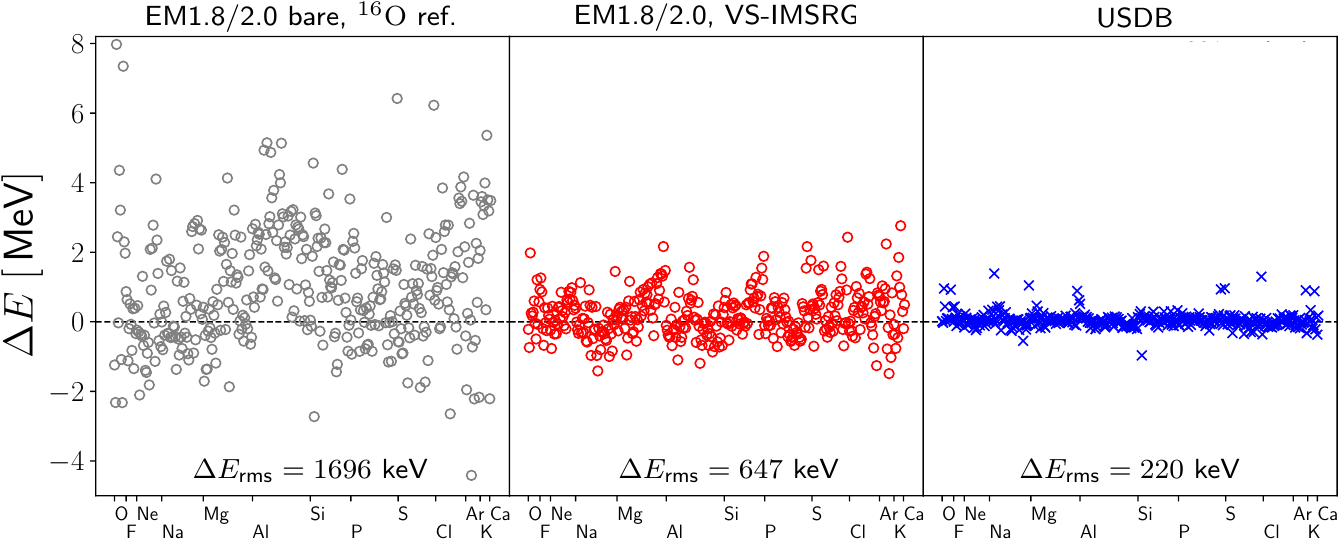}
  \end{center}
  \caption{
    Deviations between theoretical and experimental excitation energies of 391 
    $sd$-shell states, for \textbf{(a)} the EM1.8/2.0 interaction without 
    valence decoupling, \textbf{(b)} the same interaction transformed with VS-IMSRG 
    , and \textbf{(c)} the USDB interaction \cite{Brown:2006fk}. The points correspond
    to the respective root-mean-square deviations for each interaction. Figure courtesy of R.~Stroberg.
  }
  \label{fig:sm_lore}
\end{figure}

Figure~\ref{fig:sm_lore} illustrates the effect of the discussed transformations via the 
deviations between the computed and experimental energies of close to 400 levels in the
$sd$-shell. Since the EM1.8/2.0 interaction used in these calculations has a low resolution 
scale, simply using the valence-space matrix elements of the input Hamiltonian without any 
further valence-space decoupling yields a root-mean-square (rms) deviation of ``only'' about 
$1.7\,\MeV$, which is not outright disastrous. When we apply the VS-IMRSG to decouple the valence
space, the newly evolved interaction yields a much improved rms deviation of approximately
$650\,\keV$, which is better than for some of the older $sd$-shell interactions, albeit
not as good as the USDB Hamiltonian, which is shown for comparison \cite{Brown:2006fk,Magilligan:2020ux}.
This is not really surprising: USDB essentially represents the best possible fit to 
experimental data under the model assumptions, i.e., the choice of a pure $sd$-shell valence 
space, the restriction to a two-body Hamiltonian, the omission of isospin-breaking effects
from the Coulomb interaction and the nuclear interactions, and the empirical $A$-dependence
multiplying the two-body matrix elements (TBMEs). The accuracy of the VS-IMSRG results, on the
other hand, is affected by possible deficiencies in the input Hamiltonian and the use of 
the VS-IMSRG(2) truncation. Naturally, both of these aspects will be improved 
systematically in future calculations.

Phenomenological adjustments of effective Shell Model interactions like the
$A$-dependent scaling factors in the USD Hamiltonians or Zuker's monopole shift 
\cite{Zuker:2003uf} are typically attributed to the changes in the nuclear 
mean field away from the core, as well as missing three-body interactions. 
In Ref.~\cite{Stroberg:2019th}, the VS-IMSRG is used to \emph{demonstrate} 
that this is indeed the case. As described in Section \ref{sec:imsrg}, upon normal 
ordering, the three-body force gives contributions to operators of equal and 
lower particle rank, which in the Shell Model case amounts to the core energy, 
single-particle energy, and two-body matrix elements. All of these contributions 
become $A$-dependent in the VS-IMSRG, but one can shift the $A$-dependent parts 
completely into the TBMEs, like in phenomenological interactions, without changing the 
Hamiltonian matrix in the many-body Hilbert space or its eigenvalues.

Procedures like the VS-IMSRG decoupling also let us track in detail how 
operators besides the nuclear interactions evolve when they are subject
to the valence-decoupling transformation. Recall from the discussion in 
Section \ref{sec:obs} that this can even \emph{quantitatively} explain the
quenching of the Gamow-Teller strength in phenomenological Shell Model
calculations, provided two-body current contributions to the initial 
transition operator are taken into account as well. For electromagnetic
transitions, the renormalization of the one-body transition operator and
the appearance of induced terms generate at least some part of the usual
phenomenological effective charges, but a more complete treatment of 
nuclear collectivity (cf.~Section \ref{sec:imsrg})
as well the inclusion of current contributions to these operators are 
developments that need to be undertaken in the coming years.

\begin{figure}[t]
  \setlength{\unitlength}{\textwidth}
  \begin{center}
    \includegraphics[width=\unitlength]{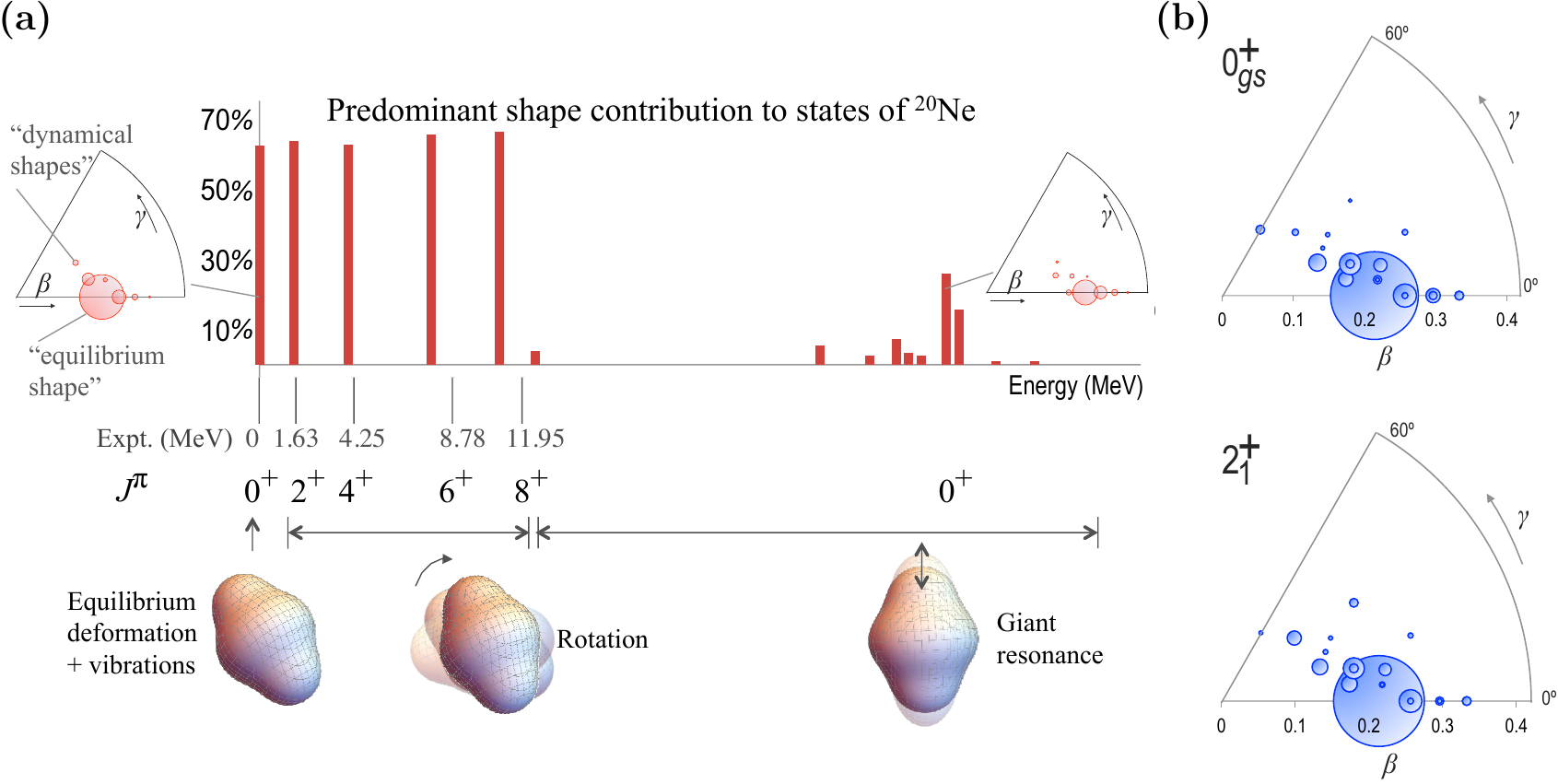}
  \end{center}
  \caption{
    SA-NCSM results for $\nuc{Ne}{20}$ in an SU(3)-adapted basis, using the 
    two-nucleon interaction \cite{Ekstrom:2013uq}. 
    Panel \textbf{(a)}: Excitation energies (horizontal axis) of the ground-state rotational 
    band ($J^\pi=0^+$ through $8^+$) and $0^+$ states, and the dominant shape
    in each state (vertical axis), indicated using the \emph{ab initio} one-body density 
    profiles in the intrinsic (body-fixed) frame. 
    Panel \textbf{(b)}: Distribution of the equilibrium shapes that contribute to the ground
    state and first excited $2^+$ state, indicated by the average deformation parameters
    ($\beta, \gamma$). See Ref.~\cite{Dytrych:2020db} 
    for additional details. Figure reprinted with permission from the American Physical Society.   
  }
  \label{fig:shapes}   
\end{figure}

\noindent\textbf{Emergence of Collectivity.} Both NCCI and VS-IMSRG
calculations with chiral NN+3N interactions have demonstrated that 
these interactions do indeed produce the telltale features of collective
behavior in nuclear spectra \cite{Bogner:2014tg,Hagen:2016rb,Caprio:2013nr,Caprio:2015wt,Caprio:2020aa}.
Upon a bit of reflection, it is not surprising that reasonable results 
on rotational bands, for instance, should be found in these approaches:
While they rely on particle-hole type expansions, the exact diagonalization
is done in a complete model space of up to $A_v$h$A_v$p excitations, where
$A_v$ is the number of valence nucleons. In contrast, euation-of-motion 
methods that typically employ 1p1h or 2p2h truncations struggle with the description of collectivity
in low-lying states \cite{Jansen:2013zr,Parzuchowski:2017yq,Morris:2018hw}, but 
they do work reasonably well for giant resonances \cite{Bacca:2013lh,Bacca:2014ye}.

As argued in Sections \ref{sec:imsrg} and \ref{sec:ncsm}, bases built on particle-hole
type expansions are not ideally suited to the description of collective correlations.
The SA-NCSM \cite{Launey:2016ef} instead uses irreducible representations of SU(3) 
or Sp(3,$\mathbbm{R}$), the dynamical symmetry groups of collective models \cite{Rowe:2016bs}, 
to achieve a much more efficient description of collective behavior in nuclei. This 
is illustrated for the case of $\nuc{Ne}{20}$ in Fig.~\ref{fig:shapes}. The SA-NCSM 
calculations \cite{Dytrych:2020db} based on the two-nucleon \NNLO\textsubscript{opt} 
potential \cite{Ekstrom:2013uq} describe the ground-state rotational band extremely 
well, all the way to the $J=8^+$ state. It is dominated by a single SU(3) irrep, associated 
with the axially elongated shape of the computed intrinsic density profile that is 
also shown in the figure.

\section{The Future to Be Written: A Look at the Challenges Ahead}
\label{sec:future}


\subsection{Rethinking the Many-Body Expansion}
\label{sec:beyond_ph}

A substantial part of the appeal of methods like CC, IMSRG and SCGF is their 
polynomial scaling. For the purposes of uncertainty quantification (UQ), we need
to be able to evaluate at least two consecutive truncation levels to assess
the convergence of the many-body expansion in nuclei for which exact calculations
are not feasible. Efforts in that direction have been in progress for some time, 
and while some methods are at a more advanced stage than others, the improved
truncations should be available for regular use within the next couple of years
\cite{Hagen:2014ve,Binder:2013fk,Miorelli:2018pb,Morris:2016xp,Hergert:2018th,Raimondi:2018qz,Soma:2020fj}.
In part, this is owing to the development of computer tools that automate tasks like
diagrammatic evaluation or angular momentum coupling \cite{Arthuis:2019vf,Tichai:2020xr}.
The computational scaling of these approaches will be of order $\mathcal{O}(N^8)$ 
or $\mathcal{O}(N^9)$, which makes applications a task for leadership-class 
computing resources for the foreseeable future. It is clear that it will 
not be feasible to just push the calculations further, since
we would then face a (naive) $\mathcal{O}(N^{12})$ scaling.

Applications where we would expect to need high-order truncations involve
nuclear states with strong collective correlations, provided we work from a
spherical reference state. As explained in Section \ref{sec:many_body}, this issue 
can likely be addressed either by using mean-field reference states with spontaneously 
broken symmetries (cf.~Section \ref{sec:cc}) or using correlated reference states
in the first place (cf.~Section \ref{sec:imsrg}), and the first applications of the
IM-GCM give credence to that idea. Moreover, there is first evidence that
the CC and IMSRG truncations converge much more rapidly for observables that
are sensitive to collectivity \cite{Novario:2020ol}, i.e., the current
state-of-the-art truncations may be sufficiently precise.

The IMSRG framework also offers perspectives for the construction of further
IMSRG hybrid methods (cf.~Section \ref{sec:imsrg}). Based on the successes of both 
the IM-NCSM and IM-GCM it would be worthwhile to use IMSRG-evolved Hamiltonians
in the SA-NCSM or techniques like the Density Matrix Renormalization Group,
which is also capable of efficiently describing strong collective correlations 
under certain conditions \cite{Rotureau:2009ml,Legeza:2015la}.

\subsection{Leveraging Computational and Algorithmic Advances}
\label{sec:smart_comp}

The progress in \emph{ab initio} many-body calculations is not simply due to
the availability of increasingly powerful computational resources, but also 
due to dedicated collaborations with computer scientists to ensure that the
available high-performance computers are used efficiently. Such collaborations
will only grow more important as hardware architectures change rapidly and
a growing demand for computing time requires users to demonstrate sufficient
efficiency to be granted access to supercomputers.

Measures to boost the numerical efficiency can also be taken at the many-body 
theory level. Efficient calculations rely on finding optimal representations 
of the relevant physical information that is encoded in the Hamiltonian. 
Algorithmic gains are possible whenever there is a mismatch, either because we 
made convenient choices, e.g., by expanding many-body states in terms of simple 
Slater determinants, or because we were not able to recognize simplifications 
beforehand, e.g., due to hidden or dynamical symmetries. 

The SRG has played a key role in addressing the first points at the level of 
the nuclear interaction over the past two decades, and SRG and IMSRG can 
be applied in novel ways to explore dynamical symmetries \cite{Johnson:2020kk}.
In the construction of a configuration space, the selection of the single-particle 
basis leaves room for optimization. 
Indeed, the natural orbitals introduced in Ref.~\cite{Tichai:2019to} lead to 
faster model-space convergence in NCSM and CC calculations, implying a more
compact Hamiltonian matrix in natural orbital representation. The efficiency
of this representation can be leveraged further by making robust importance
truncations based on analytical measures, e.g., in MBPT, CC, or IT-NCSM 
\cite{Tichai:2019jo,Roth:2009eu}. 

The aforementioned steps make use of prior theoretical knowledge, e.g.,
to identify desired decoupling patterns in interactions, or define analytical
measures for the importance of basis states. If such knowledge is not available,
or we want to avoid bias, we can leverage a myriad of Principal Component
Analysis (PCA) methods to factorize interactions or intermediate quantities
in many-body calculations \cite{Tichai:2019xz,Tichai:2019jo}. This can 
potentially even give us control over the computational scaling of nuclear
many-body methods (see, e.g., \cite{Hohenstein:2012xq,Parrish:2012os,Hohenstein:2012ez,Parrish:2014bq,Parrish:2013fe}).

\begin{figure}[t]
  \setlength{\unitlength}{\textwidth}
  \begin{center}
    \includegraphics[width=\unitlength]{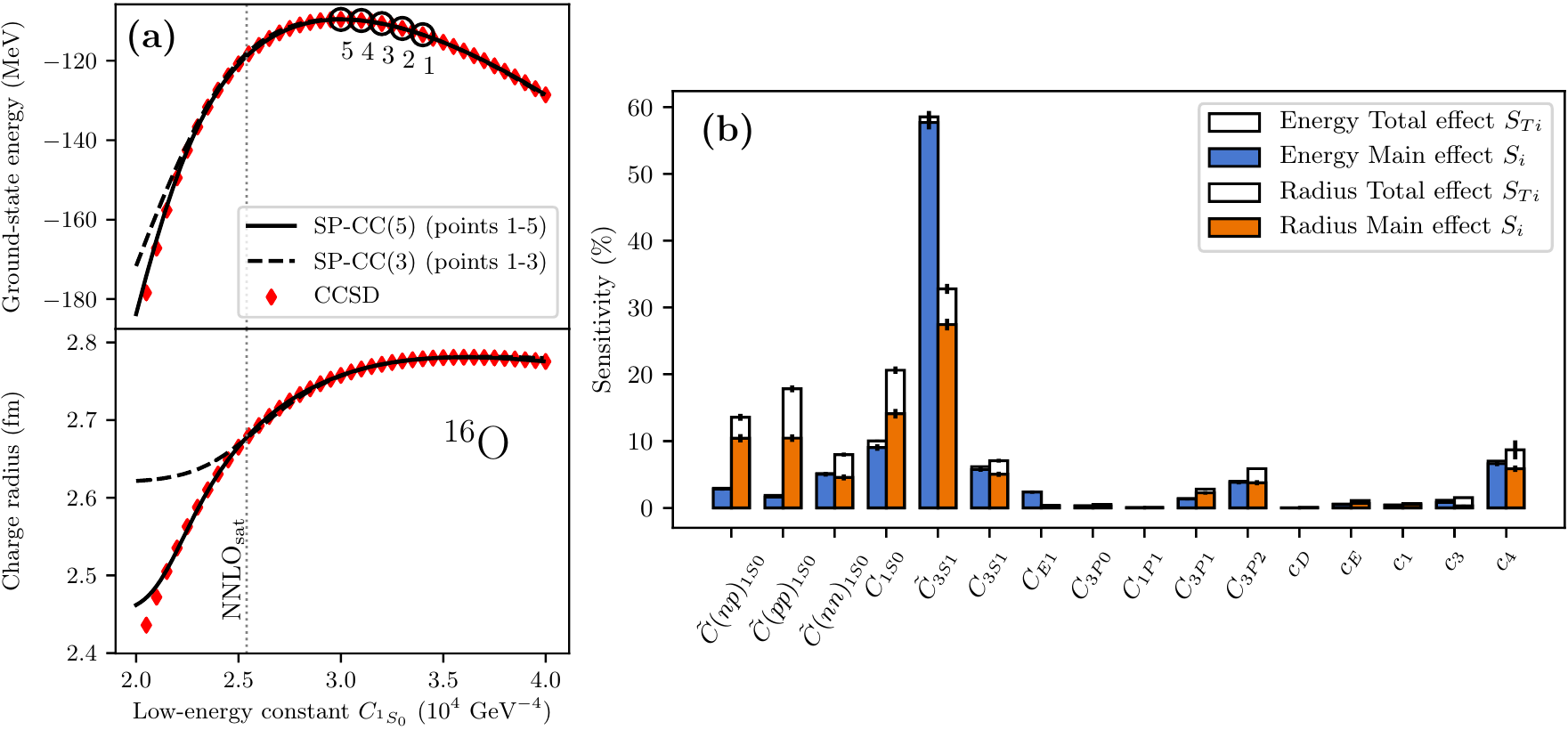}
  \end{center}
  \caption{
    Sensitivity analysis using subspace-projected CC (SPCC) method \cite{Ekstrom:2019tw}. Panel 
    \textbf{(a)} illustrates the capability of the SPCC Hamiltonian constructed from 3--5 sample
    points to predict full CCSD ground-state energies and charge radii for $\nuc{O}{16}$ over a 
    wide range of values of the chiral LEC $C_{{}^1S_0}$. Panel \textbf{(b)} shows the global 
    sensitivity of the $\nuc{O}{16}$ ground-state energy and charge radius to chiral LECs, 
    determined by evaluating over 1,000,000 quasi-MC samples from a 64-dimensional SPCC 
    Hamiltonian. Vertical bars indicate $95\%$ confidence intervals. For details, see 
    Ref.~\cite{Ekstrom:2019tw}. Figure reprinted with permission from the American Physical
    Society.
  }
  \label{fig:spcc_sensitivity}
\end{figure}

A very noteworthy development with origins in nuclear physics is Eigenvector 
Continuation (EVC) \cite{Frame:2018bv,Sarkar:2020dt}, a method for learning manifolds of
eigenvector trajectories of parameter-dependent Hamiltonians. The method
has been employed in several contexts, e.g., to stabilize high-order MBPT 
expansions \cite{Demol:2020nm} and to construct emulators for nuclear
few- and many-body calculations \cite{Konig:2019kl,Ekstrom:2019tw}. 
As an example, Fig.~\ref{fig:spcc_sensitivity} shows a global sensitivity 
analysis of CC results for $\nuc{O}{16}$ under variations of the chiral 
LECs \cite{Ekstrom:2019tw}. Eigenvector continuation was used to learn 
representations of the CCSD Hamiltonian and charge radius operators in 
a 64-dimensional subspace of the space of CCSD ground-state wave functions
for interactions with 16 varying LECs. The subspace-projected Hamiltonian 
was then sampled more than a million times on laptop, while full 
CCSD calculations of the same ensemble would be completely unfeasible.
The successful applications of EVC suggest that the method should be
further explored as a tool for improvement, emulation and UQ in other many-body 
methods in the (near) future.

\subsection{Uncertainty Quantification}
\label{sec:uq}

In typical nuclear many-body calculations as discussed in Secs. \ref{sec:ingredients} 
and \ref{sec:current} the main sources of theoretical uncertainties are the EFT
truncation of the observables and the many-body wave function, either due to
many-body expansion and/or model space truncations in configuration space 
approaches, lattice discretization effects in Lattice EFT, or the specific form of 
the wave function ansatz in QMC. If an SRG evolution is applied, there is an
additional uncertainty associated with the truncation of induced operators
(see Section \ref{sec:srg}). 

The application of Bayesian methods has led to enormous progress in the 
quantification of the EFT uncertainties \cite{Furnstahl:2014il,Furnstahl:2015sp,Wesolowski:2016ek,Melendez:2017fj,Wesolowski:2019zv,Melendez:2019ax,Drischler:2020cv}, and it would be 
highly desirable to apply the same approach to the many-body uncertainties
as well. The most challenging amongst these are the truncation of the 
many-body expansion in methods like CC, IMSRG or SCGF, and the truncation
of the free-space SRG flow of observables. In contrast, the infrared effects 
of finite-basis size truncations in HO bases --- or general orbitals that are
at some point expanded in an HO basis --- are well understood for the energy 
and other observables, and they can be systematically corrected for 
\cite{More:2013bh,Furnstahl:2014vn,Wendt:2015zl,Odell:2016qq,Forssen:2018kx}. The situation
is less clear for ultraviolet basis-size errors \cite{Konig:2014fk}, but 
this error can be suppressed by working at appropriate values of the HO
frequency.

An ideal uncertainty analysis would combine the exploration of EFT and many-body
uncertainties for nuclear observables of interest, using EC or Machine 
Learning (ML) to construct emulators that allow an efficient sampling of 
the parameter space. In such an effort, the generation of sufficient 
training data poses a significant challenge, because it would require
calculations at several truncation levels (see Section \ref{sec:beyond_ph}).
A possible strategy for mitigating this issue is to combine non-perturbative
methods with cheaper high-order MBPT in Bayesian mixed models (see
Refs.~\cite{Kejzlar:2019gc,Neufcourt:2019tx,Neufcourt:2020aq} for applications
in nuclear physics). The successful application of factorization methods 
to the nuclear many-body problem could likely resolve the issue once
and for all by reducing
the computational scaling of high-order truncations, at the cost of introducing 
an additional uncertainty from the factorization procedure.

On the road towards the destination represented by such a ``complete'' UQ 
framework, the intermediate milestones will already provide valuable insights 
into open issues in the EFTs of the strong interactions, and enable the design
of better protocols for constraining and refining EFT-based interactions and 
operators (see, e.g., Refs.~\cite{Ekstrom:2019oq,Melendez:2020zo} and references 
therein.)

\subsection{Strengthening and Employing the Hierarchy of Strong Interaction EFTs}
\label{sec:eft_future}

Strong interaction physics is a multi-scale problem, and there are good reasons
for making better use of the hierarchy of Effective (Field) Theories 
at our disposal. At the top level, we have QCD, followed by
EFTs involving hyperons that can be eliminated progressively until we 
arrive at the ``traditional'' pionful and pionless chiral EFTs (see Refs.~
\cite{Haidenbauer:2020vn,Hammer:2020lt} and references therein). At even lower 
scales, one can formulate an EFT for nuclear halos (or clusters) \cite{Hammer:2020lt} 
and make the connection to nuclear DFT and collective models, which can be 
understood as EFTs as well \cite{Furnstahl:2012fu,Furnstahl:2020pw,Papenbrock:2014kx,Papenbrock:2015ec,Coello-Perez:2015bj,Coello-Perez:2015ay,Coello-Perez:2016xi,Papenbrock:2020gd}. 

At least in principle, the different levels of this hierarchy can be connected 
either by computing observables with different theories and matching the LECs, 
or using RG flows to track in detail how the theories evolve from one into another. 
While matching procedures have been applied successfully to modern EFts in 
nuclear physics \cite{Contessi:2017ik,Hagen:2013fk,Bogner:2011vn,Dyhdalo:2017zr,Navarro-Perez:2018db,Zhang:2018ou} 
as well as efforts to match more traditional models of nuclear structure to 
\emph{ab initio} calculations \cite{Pudliner:1996cf,Duguet:2008uv,Shen:2019jv},
making the connection through RG methods is a more daunting task. While I must
admit to bias in this regard, I still consider this an effort worth 
undertaking. The success of SRG techniques in nuclear physics demonstrate how 
these methods reveal the most effective degrees of freedom even in
situations were the separation of scales is not perfectly clear. Moreover,
RGs would also reveal unexpected features of the power counting schemes, 
like the enhancement or inadvertent omission of certain operators (see Ref.~
\cite{Kolck:2020dn} and references therein). 

\noindent\textbf{Tackling Power Counting Issues.} Throughout this work, I have 
alluded to shortcomings
and issues of the current generation of chiral interactions, like the struggle
to achieve a good simultaneous description of nuclear binding energies and radii 
(see Section \ref{sec:benchmark}). 
Recent efforts to construct new, accurate chiral interactions have revealed that
this issue is connected to the use of local or nonlocal regulators, with the 
latter being favored for better descriptions \cite{Huther:2019yz,Soma:2020dw}. In another exploration
of nonlocally regularized chiral forces \cite{Drischler:2019qf,Hoppe:2019th}, 
a tension between the simultaneous description of nuclear matter and finite
was observed in the attempts to fit the chiral LECs. In QMC calculations,
it was demonstrated that the use of local regulators breaks the equivalence
of parameterizations of the interaction that are connected by Fierz identities,
in certain cases with disastrous consequences \cite{Lonardoni:2018sz}. 
Meanwhile, Epelbaum and co-workers have proposed the use of a more nuanced 
semilocal regularization scheme that applies local regulators to the long-range 
pion exchange and nonlocal regulators to the short-range contact terms 
\cite{Reinert:2018uq,Epelbaum:2020rr}. While physical arguments can be made
in favor of different regularization schemes, perhaps especially the semilocal
one, the significant scheme dependence is at odds with the principles of 
EFT, which would predict regulator artifacts to be pushed beyond the
order at which one currently works.

It has also been suggested that the scales of the chiral EFT interaction and 
the inherent scales of the many-body configuration space (e.g., IR and UV cutoffs inherited
from a HO basis, see Section \ref{sec:uq}) or coordinate space wave functions 
should not be treated independently, and that by doing so, current many-body 
approaches might at least contribute to power counting issues. There have 
been a few efforts to explore this problem, but more work is clearly required
\cite{Yang:2016sf,Yang:2020zm,Binder:2016ph,Bansal:2018vn,McElvain:2019if,Drissi:2020ya}.

\noindent\textbf{Application Needs.} Aside from the formal need to make progress
on the power counting issue,
there are also concrete application needs that call for a tighter coupling
between QCD and the nuclear EFTs. For example, the chiral EFT transition 
operator for neutrinoless double-beta decay (see Section \ref{sec:obs}) contains 
counter terms whose LECs can only be determined from Lattice QCD 
\cite{Cirigliano:2018dq,Cirigliano:2018ng,Cirigliano:2018qz,Cirigliano:2019dt}.


The dawning of a new age in our understanding of neutron stars, heralded by
the detection of gravitational waves from the neutron-star merger GW170817, 
has taken the demand for accurate neutron and nuclear matter equations of 
state to a new level (see, e.g., Ref.~\cite{Tews:2020hl} and references 
therein). While \emph{ab initio} calculations of infinite matter up to 
the saturation region based on chiral interactions are reasonably well controlled
\cite{Drischler:2020cv,Drischler:2020ik,Lonardoni:2020yo,Tews:2020hl}, 
the supranuclear densities probed in merger events are beyond the range
of validity of regular pionful chiral EFT. To increase its validity,
hyperons must be taken into account as dynamical degrees of freedom
(see \cite{Haidenbauer:2020vn} and references therein), and the entire set 
of nuclear and hyperon LECs must be readjusted at the increased breakdown 
scale. For the NN sector, this is unproblematic due to the plethora of
available scattering data. Since no direct experiments on three-neutron 
or three-proton systems are feasible, the only available experimental
constraints come from finite nuclei, which implies that the corresponding 
channels of the 3N interaction are only constrained at sub-saturation densities.
The world database of hyperon-nucleon scattering data is also quite limited.
Thus, a high-precision interaction for describing the equation of state 
at high density can only be constructed with the help of Lattice QCD
constraints on the 3N and hypernucleon LECs.

\subsection{Interfacing with Reaction Theory}
\label{sec:dynamics}

The final important research direction for the coming decade I want to discuss
here are efforts to interface the advanced \emph{ab initio} nuclear structure 
methods at our disposal with reaction theory \cite{Johnson:2019ws}.

As discussed in Sections \ref{sec:ncsm} and \ref{sec:response_and_scattering},
the NCSMC has been applied with great success to the reactions of light nuclei
at low energies, but its computational complexity makes this approach unfeasible
for nuclei beyond $A\approx 10-20$. Work has begun on a similar approach that
combines SA-NCSM with the RGM, leveraging the efficiency of the symmetry-adapted
basis to reach medium-mass nuclei \cite{Mercenne:2019vp} (cf.~Sections \ref{sec:ncsm} 
and \ref{sec:lore}). Since the RGM is just a special case of a Generator
Coordinate Method, the IM-GCM discussed in Sections \ref{sec:imsrg} and \ref{sec:obs}
is a promising candidate for extending this type of reaction calculations to 
even heavier nuclei. 

Methods that are similar in spirit to these combinations of structure approaches witht the 
RGM are the APM, which can provide an interface between structure and scattering 
in Lattice EFT (cf.~Sections \ref{sec:left} and \ref{sec:response_and_scattering}), 
as well as the GSM Coupled Channel (GSM-CC) approach, which was developed to describe 
reactions between light projectiles and targets that are treated in the GSM with 
a core \cite{Jaganathen:2014zp,Wang:2019wj,Mercenne:2019cl}.
Thus far, applications of the GSM-CC have been based on phenomenological 
valence-space interactions, but new efforts are underway to directly construct
suitable Hamiltonians based on EFT principles \cite{Fossez:2018ec,Huth:2018si}, or 
derive the effective interactions from chiral forces
with the techniques discussed in Section \ref{sec:many_body} (see \cite{Sun:2017zm,Hu:2020rw}).
Of course, the GSM-CC ideas could also be applied to the No-Core GSM 
\cite{Papadimitriou:2013dt,Shin:2017ef,Fossez:2017ty}, although the computational
complexity would limit such an approach to light nuclei.

A complementary strategy for bridging nuclear structure and reactions for
medium-mass nuclei is the construction of optical potentials for use in 
traditional reaction calculations.
In SCGF theory, the optical potential for elastic nucleon-nucleus scattering
is given by the one-body self energy, which is obtained as a byproduct
of a nuclear structure calculation, and can be used with little effort in
reaction codes \cite{Idini:2019gl}. Similarly, Rotureau et al. constructed 
optical potentials by extracting the self-energy from the Coupled Cluster
Green's Function ~\cite{Rotureau:2017km,Rotureau:2018ck,Rotureau:2020fd}. 
One can roughly view this procedure as performing a GF calculation with 
the similarity-transformed CC Hamiltonian, which does not require 
self-consistent iterations because of the CC decoupling (cf.~Section \ref{sec:cc}).
Optical potentials can also be constructed by folding scattering $T$-matrices 
with \emph{ab initio} density matrices. This technique was applied for
NCSM density matrices by two collaborations in Refs.~\cite{Burrows:2018by,Burrows:2019pj} 
and \cite{Gennari:2018nt,Gennari:2019jw}, respectively, and more applications
are underway.

While the published results from the optical-potential based approaches
are promising, an important aspect of these calculations \emph{must} be 
checked carefully in the near term: The optical potential depends on the 
resolution scale of the used chiral interactions, and the calculation 
scheme, which encompasses the truncations in the operators and many-body
method, as well as the choice of regulator in the interaction 
\cite{Furnstahl:2013zt,Duguet:2015lq}. To produce scale- and scheme-independent
observables, these choices must be matched by the reaction theory. Matching
the resolution scales is probably the easier of the two checks, but it will
require the analysis of free-space SRG transformations on the reaction theory
side. Once structure and theory are defined at a matching resolution scale,
any residual scheme dependence of the observables will give rise to the 
remaining theoretical uncertainty of the combined calculation.

\section{Epilogue}

Thus concludes our little excursion through the landscape of state-of-the-art
\emph{ab initio} nuclear many-body theory, but of course, the road goes ever 
on. I hope that this guided tour has contributed to your appreciation of the 
immense progress the community has made in the last ten years, as well as the challenges that we
are facing on the next stage of the road. None of the obstacles in our path
are unsurmountable, and while we chip away at them, results from \emph{ab initio}
calculations can make meaningful contributions to the analysis and planning
of nuclear physics and fundamental symmetry experiments. With new facilities
launching in the next couple of years, the fun will begin in earnest, so 
here's looking forward to the next decade! 

\section*{Acknowledgments}

This work has been shaped by an enormous amount of discussions over the
past decade, and naming all discussion partners would likely require 
multiple pages --- it is safe to say that if your work is cited here, we 
have likely talked in person at some point. I am deeply grateful for these 
conversations, and the spirit of the \emph{ab initio} nuclear structure and 
reactions community that fosters such exchanges. 

I would like to thank the current (and former) members of my research group 
as well as colleagues at NSCL/FRIB, who had the most immediate impact on this 
work by virtue of being a short walk away. Particular thanks go out to S.~Bogner, 
K.~Fossez, M.~Hjorth-Jensen, R.~Wirth, and J.~M.~Yao. Special thanks are owed 
to R.~Stroberg and S.~Elhatisari, who helped out with sudden requests for 
figures.

I am also grateful to the Institute of Nuclear Theory for its hospitality,
which was the venue for many of the aforementioned discussions, most recently 
during the program INT-19-2a, ``Nuclear Structure at the Crossroads''. 

The preparation of this work has been supported by the U.S. Department of 
Energy, Office of Science, Office of Nuclear Physics under Awards No. 
DE-SC0017887, DE-SC0018083 (NUCLEI2 SciDAC-4 Collaboration), and DE-SC0015376 
(DBD Topical Theory Collaboration).

\appendix
\section*{List of Acronyms}
\begin{longtable}{ll}
ADC & Algebraic Diagrammatic Construction (for Self-Consistent Green's Functions)\\
AFDMC & Auxiliary Field Diffusion Monte Carlo \\
APM & Adiabatic Projection Method (in Lattice EFT)\\
BMBPT & Bogoliubov Many-Body Perturbation Theory\\
CI & Configuration Interaction\\
CC & Coupled Cluster \\
CCSD & Coupled Cluster with Singles and Doubles excitations\\
CCSDT & Coupled Cluster with Singles, Doubles and Triples excitations\\
CCSD(T) & Coupled Cluster with Singles, Doubles and perturbative Triples excitations\\
$\chi$EFT & Chiral Effective Field Theory\\
DFT & Density Functional Theory\\
EVC & Eigenvector Continuation\\
EDF & Energy Density Functional\\
EFT & Effective Field Theory\\
GCM & Generator Coordinate Method\\
GFMC & Green's Function Monte Carlo\\
GHW & Griffin-Hill-Wheeler (equation)\\
HF & Hartree-Fock\\
HFB & Hartree-Fock-Bogoliubov\\
IM-GCM & In-Medium Generator Coordinate Method (a combination of IMSRG and GCM)\\
IM-NCSM & In-Medium No-Core Shell Model (a combination of IMSRG and NCSM)\\
IMSRG & In-Medium Similarity Renormalization Group\\
LEFT & Lattice Effective Field Theory\\
LO & Leading Order (Effective Field Theory)\\
MBPT & Many-Body Perturbation Theory\\
MR-IMSRG & Multi-Reference In-Medium Similarity Renormalization Group\\
NCCI & No-Core Configuration Interaction\\
NCSM & No-Core Shell Model\\
NCSMC & No-Core Shell Model with Continuum\\
NLO & Next-to-Leading Order (EFT) \\
\NNLO & Next-to-Next-to-Leading Order (EFT)\\ 
\NNNLO & Next-to-Next-to-Next-to-Leading Order (EFT)\\ 
\NNNNLO & Next-to-Next-to-Next-to-Next-to-Leading Order (EFT)\\ 
QCD & Quantum Chromodynamics \\
QMC & Quantum Monte Carlo\\
RG & Renormalization Group\\
RGM & Resonating Group Method\\
SCGF & Self-Consistent Green's Functions\\
SRG & Similarity Renormalization Group\\
TBME & two-body matrix elements (typically in the discussion of Shell Model interactions)\\
UCC & Unitary Coupled Cluster\\
UMOA & Unitary Model Operator Approach\\
UQ & Uncertainty Quantification\\
VMC & Variational Monte Carlo\\
VS-IMSRG & Valence-Space In-Medium Similarity Renormalization Group
\end{longtable}

\bibliographystyle{frontiersinHLTH_FPHY}
\bibliography{2019_frontiers}

\end{document}